\documentclass[]{aastex631}

\shorttitle{DTW as a means to assess solar wind time series}
\shortauthors{Samara et al.}

\usepackage{subfigure}

\hypersetup{colorlinks=true,citecolor=cyan,urlcolor=cyan,linkcolor=blue}

\begin{document}

\title{The Dynamic Time Warping as a Means to Assess Solar Wind Time Series}

\correspondingauthor{Evangelia Samara}
\email{evangelia.samara@kuleuven.be}

\author[0000-0002-7676-9364]{E. Samara}
\affiliation{SIDC, Royal Observatory of Belgium, Brussels, Belgium and CmPA/Dept.\ of Mathematics, KU Leuven, Leuven, Belgium \\}

\author{B. Laperre}
\affiliation{CmPA/Dept.\ of Mathematics, KU Leuven, Leuven, Belgium}

\author{R. Kieokaew}
\affiliation{IRAP, University of Toulouse, Toulouse, France}

\author{M. Temmer}
\affiliation{University of Graz, Institute of Physics, Graz, Austria }

\author{C. Verbeke}
\affiliation{SIDC, Royal Observatory of Belgium, Brussels, Belgium and CmPA/Dept.\ of Mathematics, KU Leuven, Leuven, Belgium \\}

\author{L. Rodriguez}
\affiliation{SIDC, Royal Observatory of Belgium, Brussels, Belgium \\}

\author{J. Magdaleni\'{c}}
\affiliation{SIDC, Royal Observatory of Belgium, Brussels, Belgium and CmPA/Dept.\ of Mathematics, KU Leuven, Leuven, Belgium \\}

\author{S. Poedts}
\affiliation{CmPA/Dept.\ of Mathematics, KU Leuven, Leuven, Belgium and Institute of Physics, University of Maria Curie-Sk{\l}odowska, Lublin, Poland}

\begin{abstract}
{During the last decades, international attempts have been made to develop realistic space weather prediction tools aiming to forecast the conditions on the Sun and in the interplanetary environment. These efforts have led to the development of appropriate metrics in order to assess the performance of those tools. Metrics are necessary to validate models, compare different models and monitor improvements of a certain model over time. In this work, we introduce the Dynamic Time Warping (DTW) as an alternative way to evaluate the performance of models and, in particular, to quantify differences between observed and modeled solar wind time series. We present the advantages and drawbacks of this method as well as applications to WIND observations and EUHFORIA predictions at Earth. We show that DTW can warp sequences in time, aiming to align them with the minimum cost by using dynamic programming. It can be applied in two ways for the evaluation of modeled solar wind time series. The first, calculates the sequence similarity factor (SSF), a number that provides a quantification of how good the forecast is, compared to an ideal and a non-ideal prediction scenarios. The second way quantifies the time and amplitude differences between the points that are best matched between the two sequences. As a result, DTW can serve as a hybrid metric between continuous measurements (e.g., the correlation coefficient), and point-by-point comparisons. It is a promising technique for the assessment of solar wind profiles providing at once the most complete evaluation portrait of a model.}

\end{abstract}


\section{Introduction}

Improving the accuracy of space weather forecasting at Earth is directly linked to a better understanding of the capabilities provided by the space weather prediction models, currently available to the operational and scientific community. A large number of such models have been developed over the past decades, aiming to reconstruct the solar corona and the heliospheric environment \cite[see e.g.,][for an overview of the available models and their capabilities]{MacNeice2018}. Some models are fully empirical, such as the PDF model \citep{BussyVirat2014}, the PROJECTZED \citep{riley2017} and the Analogue Ensemble model \citep[AnEn;][]{Owens2017}. Others require some extensive empirical tuning such as e.g., the Wang-Sheeley-Arge model \citep[WSA;][]{ArgePizzo2000, arge03, Arge04} which is one of the most widely used operational coronal models, or the Empirical Solar Wind Forecast model \cite[ESWF;][]{reiss16}, which relates the areas of the coronal holes as observed in extreme ultraviolet (EUV) with the solar wind speed prediction at the Lagrangian point 1 (L1), based on \citet{vrsnak07}. Some models combine empirical approximations with physics \citep[see e.g., the MULTI-VP model by][]{Pinto17}, that uses the PFSS model for the reconstruction of the lower corona and relies on the solution of magnetohydrodynamic (MHD) equations for the calculation of the solar wind properties in the upper corona. Another large category of models are the physics-based models that are consisting entirely of MHD codes. Such models can either reconstruct the global solar corona alone, such as the MAS model \citep{Mikic1999, Lionello2003} or the heliospheric domain alone, such as the heliospheric parts of ENLIL \citep{odstrcil99} and EUHFORIA \citep{pomoell18}. Some models can reconstruct both domains, such as the MAS model for corona and heliosphere \citep{Riley2011} or the AWSoM model \citep{meng2015}. Last but not least are models based on tomographic techniques. For example, HelTomo \citep{Jackson1998, Jackson2002, Jackson2020} is a tool that reconstructs the solar wind including coronal mass ejections (CMEs) by employing interplanetary scintillation of astronomical radio sources that are viewed through the ambient solar wind plasma. 

Space weather prediction models are subject to continuous changes and improvements. Therefore, it is of utmost importance for space weather forecasting applications to record and quantify these changes over time. Differences in performance should also be recorded among different models, e.g., MAS vs ENLIL vs EUHFORIA, or among combinations of different models. For example, an overview of the coupling between the WSA coronal model and the time-dependent MAS solar wind model is given in \citet{Linker2016} and an overview of the coupling between MULTI-VP coronal model and EUHFORIA-heliosphere in \citet{Samara2021}. Metrics is the only way to quantitatively understand which set-up provides the best results based on the user's needs and goals every time.

A number of metrics proposed throughout the years is being used by the scientific community. \citet{Owens2018} presents an overview by categorizing them in two large groups: the \textit{point-by-point} metrics and the \textit{time-window} metrics. The former category includes the so-called error functions such as the mean-square error (MSE), the mean absolute error (MAE), the root-mean-square error (RMSE) or general skill scores constructed by the aforementioned. All of them are widely used to compare the amplitudes between observed and predicted time series in a point-by-point manner. Another sub-category of the point-by-point metrics is the binary metrics. Binary metrics show whether large-scale structures of the solar wind (e.g., high-speed streams abbreviated as ``HSSs") have arrived at the point of interest or not. This technique is based on hit/miss statistics between the observed and predicted time series once specific requirements are fulfilled (e.g., a HSS arrives within a specific time window and has an an amplitude and gradient value above a specific threshold). For example, by setting a solar wind velocity threshold at 500 km/s and by taking into account a specific interval in time (e.g., 2 days), this binary technique indicates whether a solar wind feature was captured from both the observed and predicted time series (\textit{hit}), or it was observed but not predicted (\textit{miss}), or it was predicted but not observed (\textit{false alarm}) and lastly, whether neither observations nor forecast indicated the arrival of an event \cite[\textit{true negative}, see e.g.][]{reiss16, Hinterreiter19}.  

The mentioned metrics are necessary to quantify differences between observed and predicted time series. They are not sufficient, though, to present the complete picture of the comparison between observations and predictions. The most significant deficiency is found in their weakness to quantify time-uncertainties. A solution to this problem can be given by using time-window metrics. The simplest time-window approach is to conduct a case-by-case analysis in which the time difference between the arrival/ending times of observed and modeled events can be recorded. Nevertheless, this is a time-consuming procedure that requires a priori definition of each event, as discussed in \citet[][]{Owens2018}. The same author introduces an alternative time-window approach, the so-called \textit{scale-selective approach} based on which agreement between observations and predictions is taken into account at a range of time scales. As the time scales become increasingly coarse, false alarms and missed events are canceling out. This technique allows an assessment of the time scales at which the forecast provides a specific accuracy level. 

\begin{figure}
\gridline{\fig{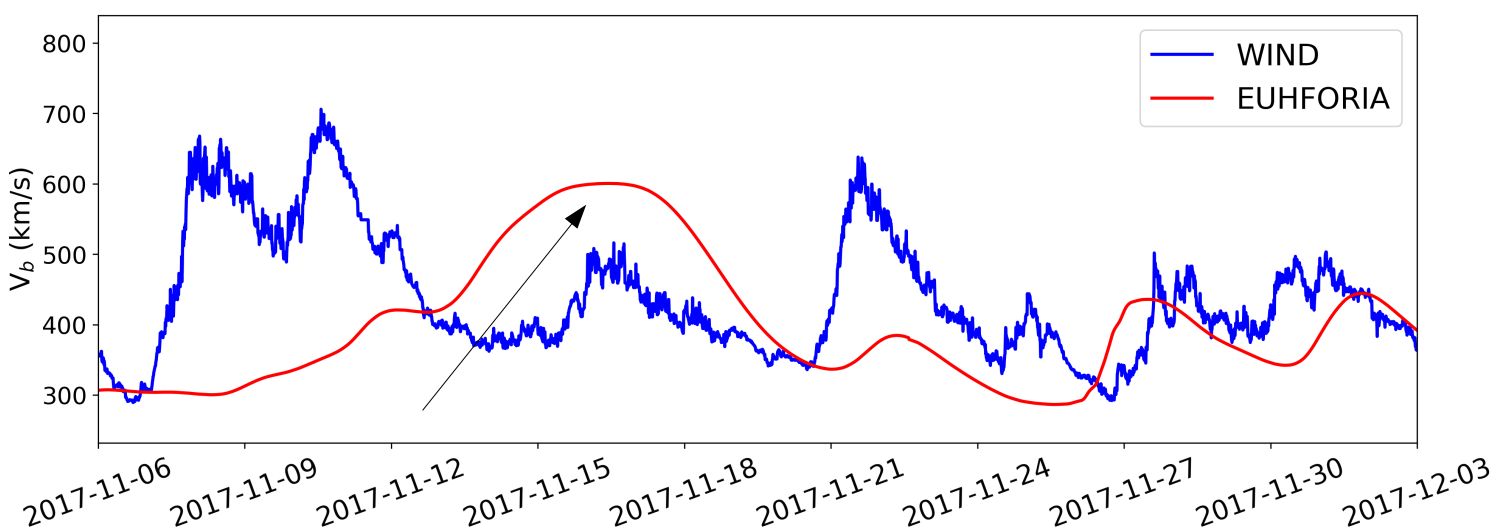}{0.75\textwidth}{(a)} }
\gridline{\fig{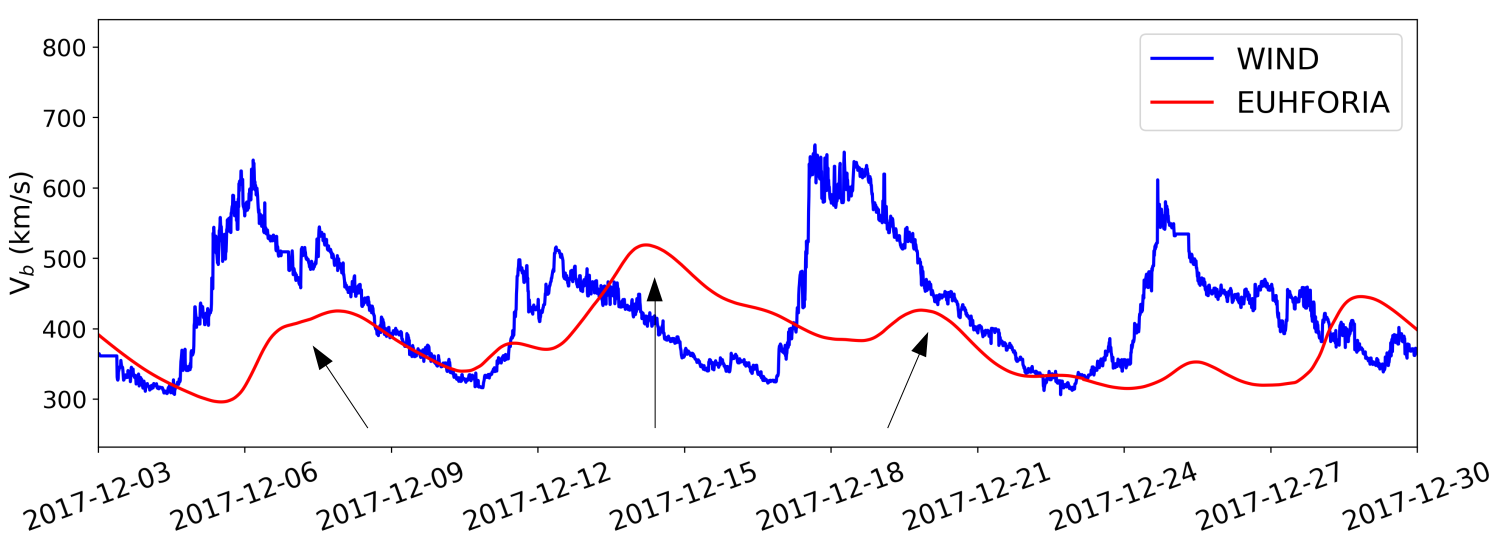}{0.75\textwidth}{(b)} }
\caption{Solar wind speed time series of observed (blue) and predicted (red) data for CR 2197 (panel a) and CR 2198 (panel b). The arrow in panel (a) points to the unclear modeled HSS structure which we are not sure if it corresponds to the observed HSS between 2017-11-06 and 2017-11-12 or to the second, slower HSS observed between 2017-11-15 and 2017-11-18. The first two arrows in panel (b) show the two HSSs that are reproduced by the model but arrive late compared to observations. The third arrow in the same panel corresponds to the uncertain HSS case for which we are not sure if it was predicted by the model.}
\label{Fig:ExamplesDTW}
\end{figure}

In this paper, we introduce the Dynamic Time Warping (DTW) technique as an additional method to assess solar wind time series. It is a powerful tool that combines the qualities of both point-by-point and time-window metrics and, as a result, it is very useful for providing a more complete assessment between predictions and observations. DTW finds matching solar wind signatures (even weak ones of slow speed, or HSSs) in order to quantify underestimated or overestimated arrival times, amplitudes and their duration. Even though this technique has been introduced many decades ago and has been extensively used by other scientific fields, we will show how it can be applied for the evaluation of solar wind time series, for better understanding of modeled solar wind profiles (long-term and short-term) and their comparison with measurements. DTW can also be applied for the evaluation of time series during CME arrivals at a specific point of interest. Nevertheless, the reasons that led us to focus on solar wind time series are summarized as follows: 

\begin{itemize}
    \item [\textbullet] The solar wind forecast at Earth (or at any other point in the heliosphere) usually deals with long and variable time series (in the order of days, weeks or even months, depending on the goals). For the correct prediction and assessment of both the fast and the slow solar wind, we need to evaluate the whole range of the available data set. This procedure is much more complicated than assessing CME signatures for which the arrival of the shock/magnetic cloud is well defined during a limited time interval (in the order of hours) and thus, allows the easy quantification of both the amplitude and the time delay. 
    \item [\textbullet] When we focus on the assessment of fast streams in the solar wind, the majority of modeled HSSs arrive later or earlier than observed \citep[see][]{Hinterreiter19}. As a result, there is always a time difference between the observed and predicted large-scale variations in the solar wind time series. DTW is the ideal technique to evaluate these variations since it aligns time series by warping them in time.
    \item [\textbullet] For some cases it is not clear how and if the predicted data should be matched with the observed data (see example in Fig.~\ref{Fig:ExamplesDTW}a). Hence, a technique that quantifies the overall performance of two sequences is required, regardless the identified structures.
\end{itemize}

\noindent Figure~\ref{Fig:ExamplesDTW} shows the comparison of the observed \cite[by WIND satellite;][]{WIND} and predicted (by EUHFORIA) solar wind time series for Carrington Rotations (CRs) 2197 and 2198. It covers the time interval from 2017-11-06 to 2017-12-30. Based on the Richardson-Cane list \citep[][]{RClist2003, RClist2010} there was only one CME recorded influencing Earth during that period. This was on 2017-12-25, so, we will not comment on the time interval after this date to avoid any uncertainties. Between 2017-11-06 and 2017-12-24, seven HSSs were identified based on the criteria proposed by \citet[][]{Jian2006}. We cross-validated the associated coronal holes on the Sun by checking the Atmospheric Imaging Assembly \cite[AIA,][]{Lemen2012} images from the Solar Dynamics Observatory \citep[SDO;][]{SDO}, as well as NOAA full-Sun drawings\footnote{https://www.ngdc.noaa.gov/stp/space-weather/solar-data/solar-imagery/composites/full-sun-drawings/boulder/}. Figure~\ref{Fig:ExamplesDTW}a shows an example of an unclear HSS reconstructed case. More specifically, it is uncertain whether the first modeled HSS (red color, see black arrow) corresponds to the first observed HSS (velocity up to $\approx$700~km/s) which impacted Earth between 2017-11-06 and 2017-11-12 or to the second, slower HSS observed between 2017-11-15 and 2017-11-18 (velocity up to $\approx$500~km/s). Therefore, instead of trying to determine whether and which HSS arrived at Earth, it is more practical to determine how similar the time series are overall. In Fig.\ref{Fig:ExamplesDTW}b we notice three HSSs influencing Earth between 2017-12-03 and 2017-12-24. The first two, were modeled later-than-expected by EUHFORIA. For the third HSS, we are uncertain if it was actually predicted by the model. These examples show that assessing the time series as a whole, instead of trying to evaluate uncertain cases one-by-one, is often the optimal solution.

\section{Benefits, drawbacks and restrictions of DTW}

DTW is a well-known technique for estimating the similarities between two time series that have a similar pattern but differ in time \citep[][and references therein]{Gorecki2013, KeoghPazzani2001}. It was initially developed for speech recognition purposes where specific words are recognized by their audio signal profiles \citep{Itakura1975, SakoeCHiba78, Myers1981, Muller2007}. Over the years, it met great interest by other scientific fields such us meteorology, robotics, medicine, music processing, manufacturing, and it is widely used in data-mining for time series clustering and classification purposes \citep[][and references therein]{KeoghPazzani98, KeoghPazzani2001}. DTW is a distance measure, similarly to the Euclidean distance. The main difference from the latter, is that DTW can manage time distortions \citep{Zhang2017} and does not always obey to the triangular inequality \cite[][]{Vidal1988}. More specifically, it allows the drifting of the vector components along the time axis when a comparison between two sequences is made. The sequences are eventually non-linearly warped along the time dimension to match each other \citep{Muller2007, Gorecki2013}. A recent study by \citet{Laperre2020} used this technique for evaluating the Dst forecast with machine learning. In this work, we apply the DTW technique for the first time as a means to quantify differences between observed and modeled time series in solar wind forecasting. For that, we use DTW in two different ways. The first way is based on the so-called \textit{cumulative cost} or \textit{DTW score}, a single number that DTW produces that translates to an estimate of the minimum ``effort" the technique put to align the observed and predicted time series of the solar wind parameters. After DTW estimates the best (and less costly) alignment between the data points, we can explicitly quantify the differences in time and amplitudes between the two sequences. This is the second way that the method can be exploited based on which, we can derive a number of relevant statistics. 

\subsection{Properties and requirements of DTW}

DTW determines how similar two time series are, by providing a temporal alignment between them, in an optimal way and under certain constraints \citep{Muller2007}. Strictly speaking, DTW is not a metric as it violates the triangular inequality\footnote{For the definition of a distance measure as a metric see \url{https://mathworld.wolfram.com/Metric.html}}. Nevertheless, it can be used as one by obeying to the following three principles \citep{Muller2007, Jeong2011}:


\begin{enumerate}
    \item The first and last point of one sequence should be matched with the first and last point of the other sequence (but it is not necessary that their matches are unique).
    \item The mapping of the elements should be monotonically increasing (it cannot go backwards in time).
    \item There should be no data gaps, namely, every point should be matched with at least one other point (continuity rule).
\end{enumerate}

\noindent The method finds the optimal alignment between two sequences by finding the path through the DTW cost matrix that minimizes the total cumulative cost among all other possible paths \citep{Ratanamahatana&Keogh, KeoghPazzani2001, Muller2007}. This path is called the \textit{warping path} and characterizes the mapping between the two time series of interest. The DTW cost matrix is filled based on the following equation: 

\begin{equation}
\centering
    D(i,j) = \delta(s_{i},q_{i})+\hbox{min}\{D(i-1, j-1), D(i-1, j), D(i, j-1)\}
\label{eq:DTW}
\end{equation}

\noindent where $D(i,j)$ is the cumulative DTW cost or distance, and $\delta(s_{i},q_{i}) = |s_{i} - q_{i}|$ corresponds to the Euclidean distance between the point $s_{i}$ from one time series and the point $q_{i}$ from the other time series. The first element of the array $D(0,0)$ is equal to $\delta(s_{0},q_{0})$. After the DTW cost matrix is filled, the warping path can be efficiently found. For that, dynamic programming is used to evaluate the recurrence that defines the cumulative DTW distance $D(i,j)$ as the Euclidean distance, $\delta$(i,j), found between two elements and the minimum of the cumulative distances of the adjacent elements \citep{Ratanamahatana&Keogh, Gorecki2013}.

\subsection{Drawbacks of DTW and ways to eliminate them}

The weak point of DTW is the so-called ``pathological alignment problem'', namely, the fact that a data point in one time series can be linked to a large subsection of points of the other time series \citep[][and references therein]{KeoghPazzani2001, Gorecki2013}. These pathological alignments are called singularities. Many techniques have been proposed to alleviate this problem. Three of the most widely used techniques can be summarized as follows \citep{KeoghPazzani2001}:

\begin{enumerate}

\item \textit{Windowing} \citep{Berndt&Clifford94}:
Allowable elements of the cumulative matrix can be restricted to those that fall into a warping window. In other words, a data-point in one time series cannot be matched with all the data-points from the second time series, but it can only be matched with the points found in a specific time window. Although this approach constrains the maximum size of a singularity, it does not prevent their occurrence.

\item \textit{Slope weighting} \citep{Kruskal&Liberman83, SakoeCHiba78}:
If the equation that calculates the accumulated cost is replaced with:
$D(i,j) = \delta(i,j) + \hbox{min}\{ D(i-1,j-1) , X D(i-1,j ) , X D(i,j-1)\} $, where $X$ is a positive real number, we can constrain the warping by changing the value of $X$. As $X$ gets larger, the warping path is increasingly biased toward the diagonal.

\item \textit{Step patterns or slope constraints} \citep{Itakura1975, Myers1981}: We can replace the cumulative cost equation with $D(i,j) = \delta(i,j) + \hbox{min}\{ D(i-1,j-1) , D(i-1,j-2) , D(i-2,j-1) \}$, which corresponds to the step-pattern. Using this equation the warping path is forced to move one diagonal step for each step parallel to an axis.
\end{enumerate}

Besides these three methods, a number of DTW variants have been proposed with the aim of reducing singularities. For example, the Derivative DTW \citep[DDTW,][]{KeoghPazzani2001}, the Weighted DTW \citep[WDTW,][]{Benedikt2008} or the Value-Derivative DTW \citep[VDDTW,][]{Kulbacki2002} are some of such techniques. The output from these three methods were also tested for the cases investigated in this work but they yielded ambiguous results. Therefore, for the purposes of this study, we only focus on the classic DTW as it leads to the most transparent output.

It is important to note that even though singularities are considered a drawback of DTW by many authors, we find that for solar wind time series, the best alignment between two sequences cannot be achieved without them. Identical time series have no singularities. As we deal with non-identical (but similar) time series, we expect singularities to occur as a way of optimally matching the points between them.

\section{Data pre-processing}
\label{section:data pre-processing}

In order to efficiently apply DTW for the evaluation of modeled solar wind time series, we need to adopt a number of constraints that will best serve our needs. We focus on the evaluation of the solar wind bulk speed as it is the best modeled solar wind characteristic in EUHFORIA \citep[v1.0.4, see e.g.,][]{pomoell18}. The same procedure can be applied for other solar wind signatures such as e.g.\ density, magnetic field, temperature etc. 

\subsection{Sensitivity on the initial and final points of the sequences}
\label{subsection: sensitivity_on_ini&final_points}

We are interested in approximately two years of continuous solar wind data (November 2017 - September 2019). The first important task is to separate this interval in smaller periods for two reasons: first, because it will be easier for future users to compare an upgraded version of EUHFORIA (or any other model) with the current version. Secondly, because it is faster to evaluate shorter periods rather than a single, extended one since DTW has a quadratic ($O(nw)$) complexity \citep[][and references therein]{KeoghPazzani2001, Ratanamahatana&Keogh}. Nevertheless, recent studies have shown that for many applications DTW complexity is reduced to linear \citep[$O(n)$,][]{TimeComplexityDTW}, similar to the complexity of the Euclidean distance.

\begin{figure}
\centering
\gridline{\fig{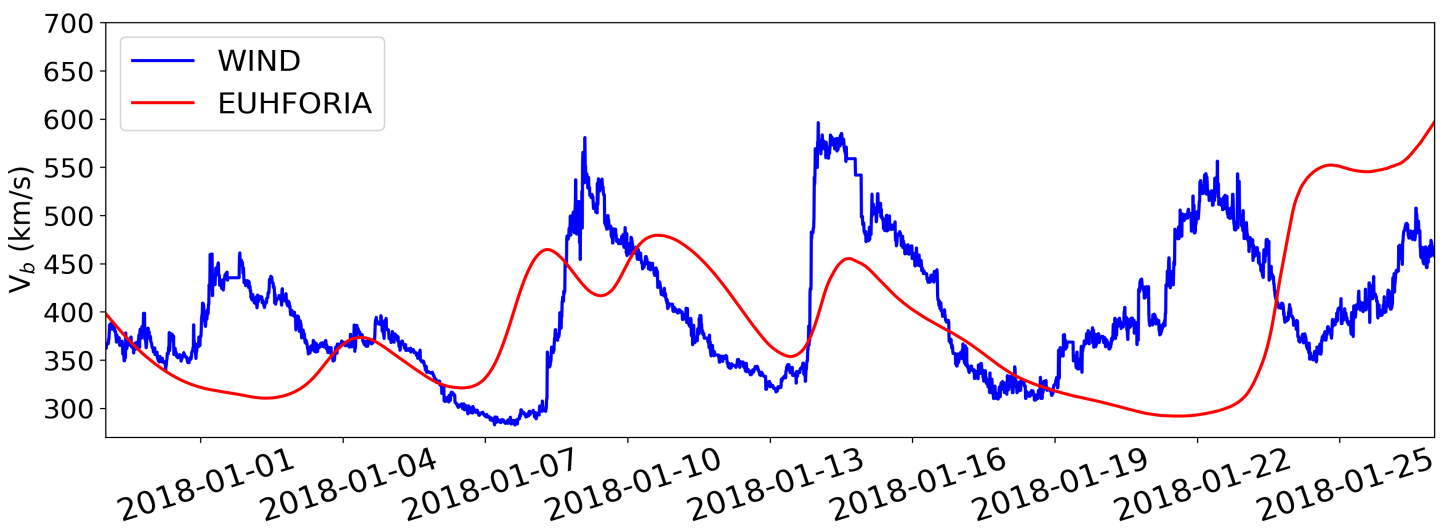}{0.70\textwidth}{(a)} }
\gridline{\fig{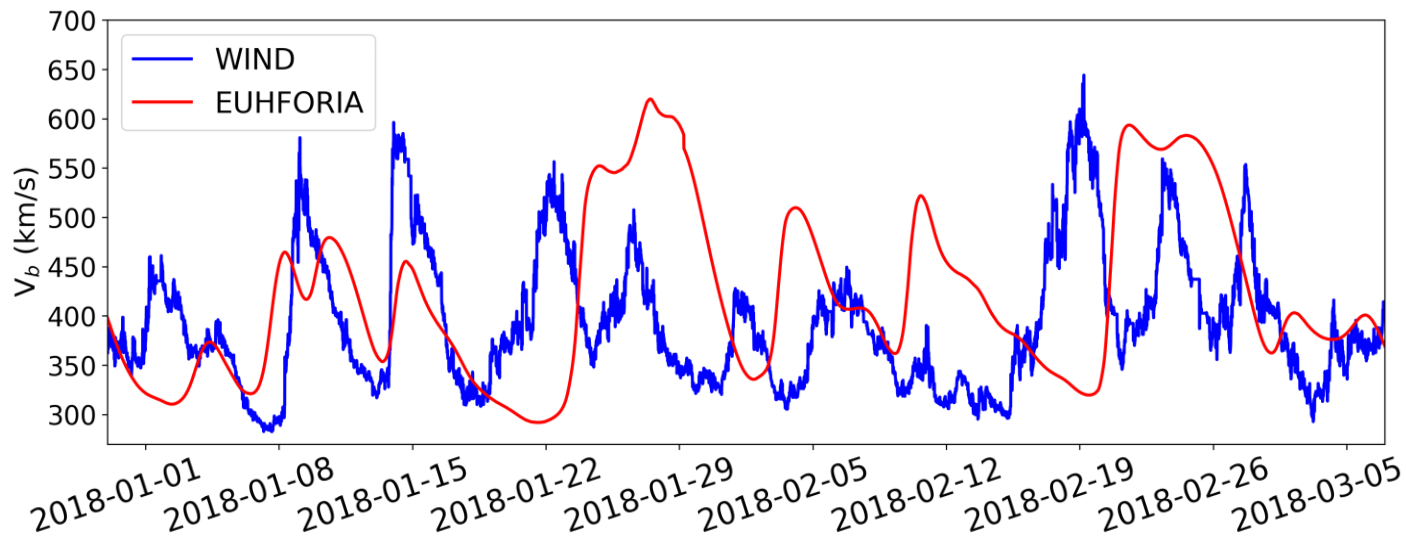}{0.7\textwidth}{(b)} }
\caption{Example of the extension of a not well selected time interval, when the influence of a HSS is still ongoing at the end of that interval. Panel (a): CR 2197 in which the effect of a HSS is still ongoing. Panel (b): extension of CR 2197 until a point in which the observed and predicted data are very close to each other during quiet solar wind levels. }
\label{Fig:CR2199-2200-2201}
\end{figure}

Special attention is needed during the division of the big time interval into smaller periods since DTW is highly affected by the initial and final points of the sequences. One way to do this, is to split the whole range to time segments of a CR. However, this is not always the best solution as a HSS may be cut by the artificial start and end time of the CR (see Fig.~\ref{Fig:CR2199-2200-2201}a). For an optimal evaluation with DTW, we require that solar wind features (such as HSSs) are fully covered, preferably with quiet background solar wind times before and after the time period under study (see comparison between Fig.~\ref{Fig:CR2199-2200-2201}a and \ref{Fig:CR2199-2200-2201}b).

\begin{figure}
\centering
\gridline{\fig{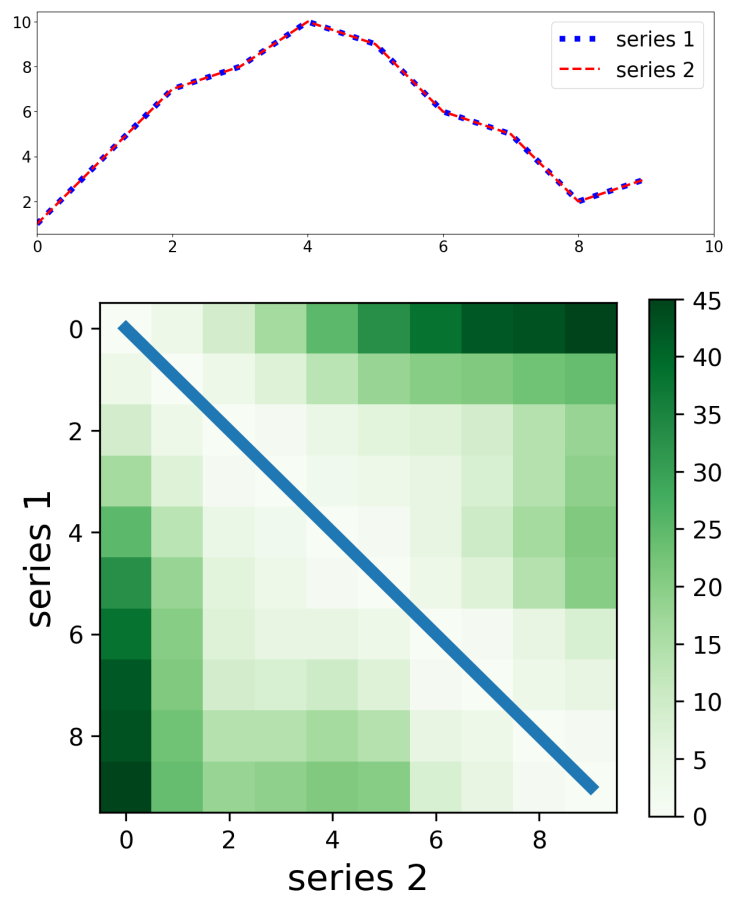}{0.4\textwidth}{(a) DTW score = 0, Sum(diag) = 0}
            \fig{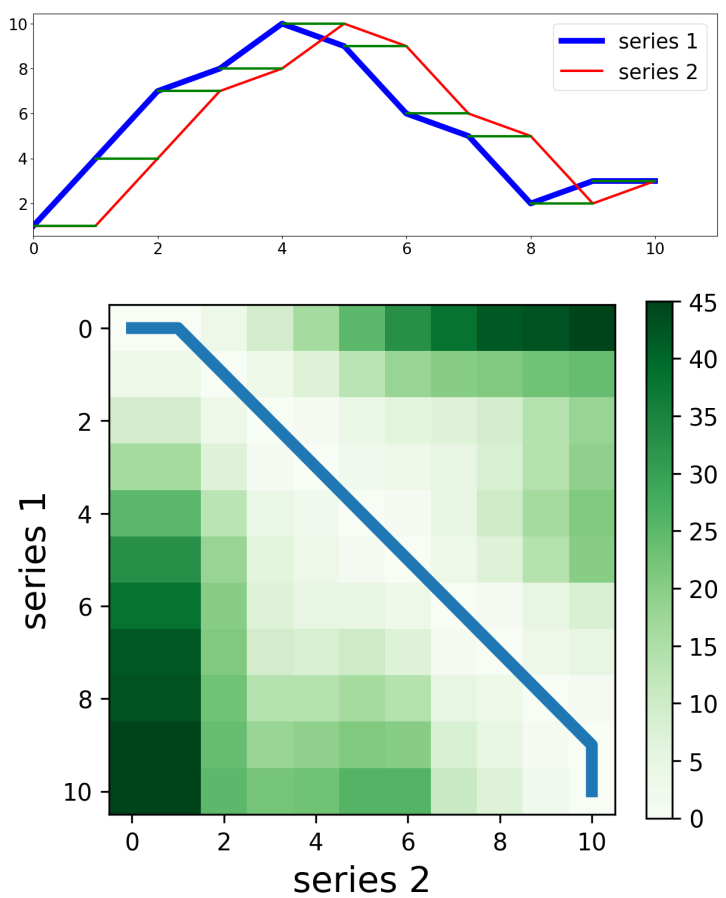}{0.4\textwidth}{(b) DTW score = 0, Sum(diag) = 18}}
            
\gridline{\fig{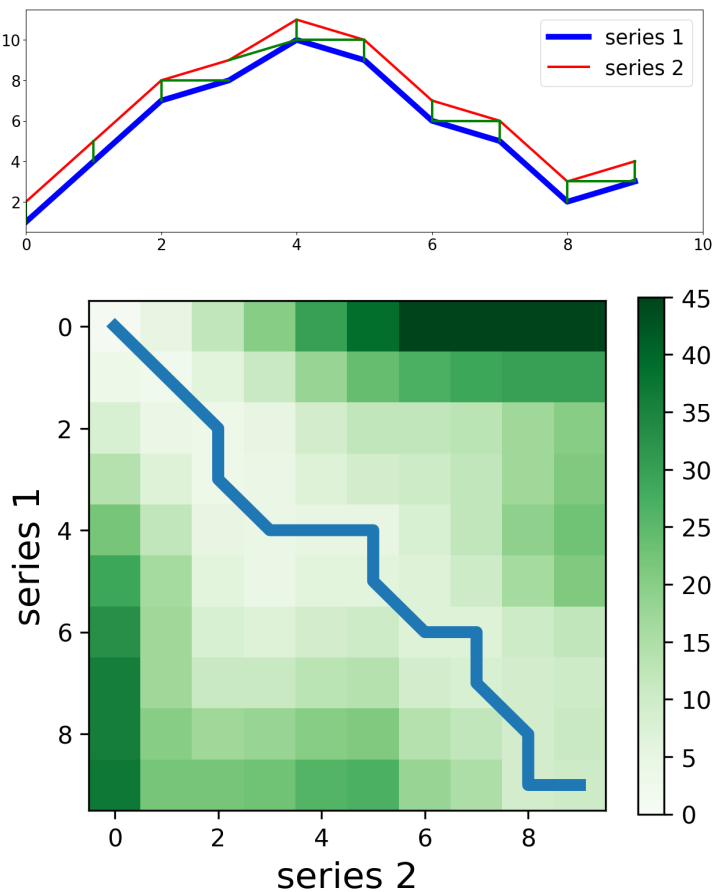}{0.4\textwidth}{(c) DTW score = 10, Sum(diag) = 55} 
            \fig{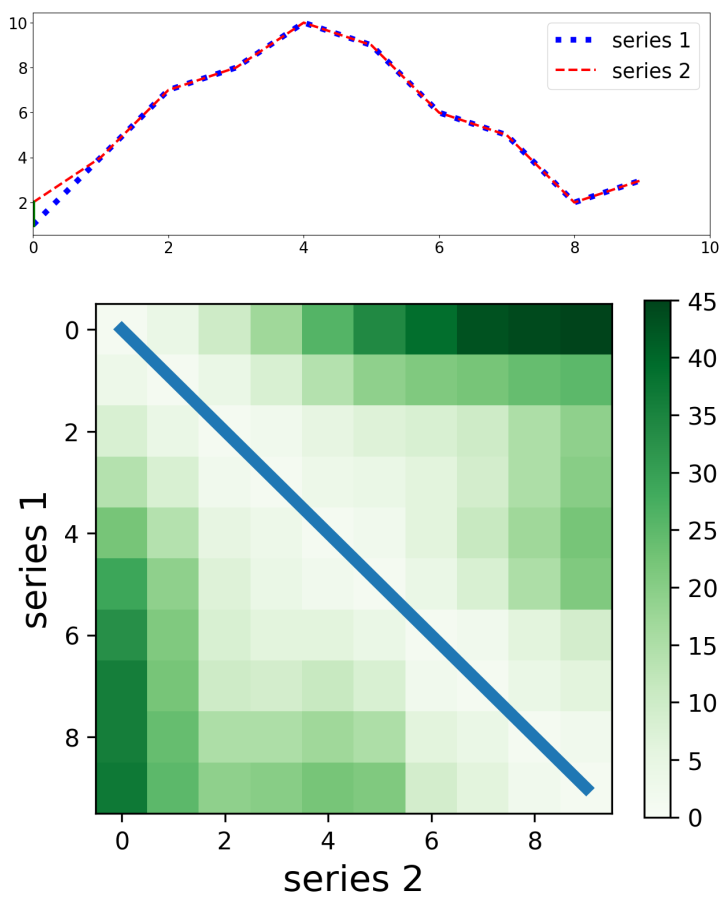}{0.4\textwidth}{(d) DTW score = 1, Sum(diag) = 10} }

\caption{Alignments, warping paths and heat-maps of random time series. The DTW score and the sum of the cost matrix diagonal are also calculated for each case. The DTW alignments are shown in green when one series is shifted compared to the other. Panel (a): the two sequences are identical. Panel (b): series 2 from panel (a) has been shifted along the $x$-axis in respect to series 1. The initial and final points are kept fixed. Panel (c): series 2 from panel (a) has been shifted along the $y$-axis in respect to series 1. Panel (d): series 2 is identical to series 1, except its initial point which has been shifted one value along the $y$-axis.  }
\label{Fig:TestCases_OtherWay_all}
\end{figure}

We now consider a simple example with random time series. Figure~\ref{Fig:TestCases_OtherWay_all} shows how DTW behaves in four different cases. In the first case, we compare two identical time series. In the second case, we shift one of the two sequences one value along the $x$-axis, keeping the same initial and final points. In the third case, we shift one of the two sequences one value along the $y$-axis, maintaining the same exact pattern. In the last case, we keep the two time series as in the initial example, but we only shift the first element of the red time series one value up along the $y$-axis. The DTW cost matrix for every case is shown as a green heat-map. The $x$- and $y$- axis of the heat-maps correspond to the index number of the elements in each time series. Darker shades of green correspond to higher estimated costs based on Eq.~\ref{eq:DTW}. Each matrix cell corresponds to a link between two points of the two series. The last (bottom right) cell in the DTW cost matrix indicates the DTW score which is a representation of the minimum effort the technique put to align the two time series. 

In the first case (Fig.~\ref{Fig:TestCases_OtherWay_all}a), where identical time series are considered, the DTW score is zero and the warping path is the diagonal of the cost matrix. This means that every element in sequence 1 was only matched with its exact corresponding (and identical) element in sequence 2. In the second example (Fig.~\ref{Fig:TestCases_OtherWay_all}b), the DTW score is again equal to zero, but the warping path deviates from the diagonal due to the time series being shifted along the x-axis. This means that the first and last points from one sequence were matched twice with points from the other sequence, thus, creating two singularity points. More specifically, the first point from series 1 was matched with two points from series 2 (horizontal shifting on the warping path) while the last point from series 2 was matched with two points from series 1 (vertical shifting in the warping path). The question is how we can distinguish between the first and second case, when no heat-maps/warping paths are provided, e.g., in large data sets, when we want to quickly evaluate the sequences of interest. The answer is given by calculating the sum of the diagonal elements of the cost matrix. For example, when we compare identical time series, the sum of the elements along the diagonal should be zero, opposite to the case where the time series are shifted along the $x$-axis.

In the third case (Fig.~\ref{Fig:TestCases_OtherWay_all}c), in which we have shifted the red sequence along the $y$-axis, we see that the DTW score and sum of the diagonal are different than zero. Moreover, the warping path deviates a lot from the diagonal. To understand better how DTW behaves during a vertical shifting of a sequence, we compare in Fig.~\ref{Fig:TestCases_OtherWay_all}d the same time series as in Fig.~\ref{Fig:TestCases_OtherWay_all}a having shifted only the first element of the red sequence along the $y$-axis. The DTW score and sum of the diagonal in this case is again different than one, but the warping path is the diagonal itself.

\begin{figure}
\centering
\gridline{\fig{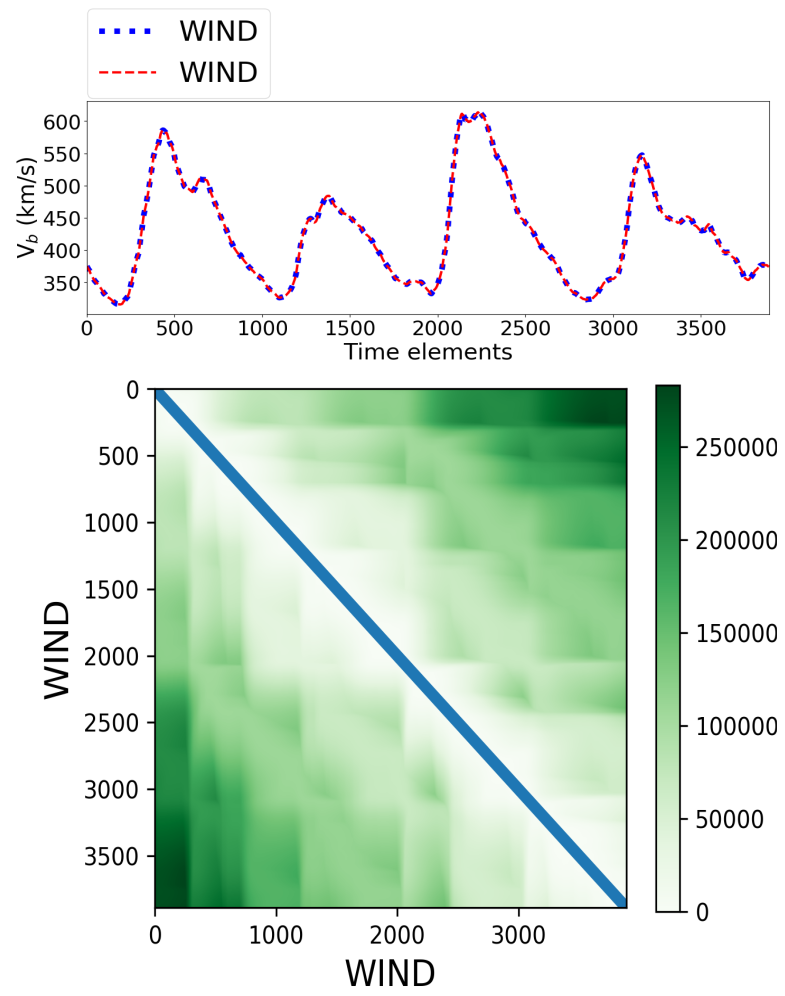}{0.305\textwidth}{(a) DTW score = 0, Sum(diag) = 0}
            \fig{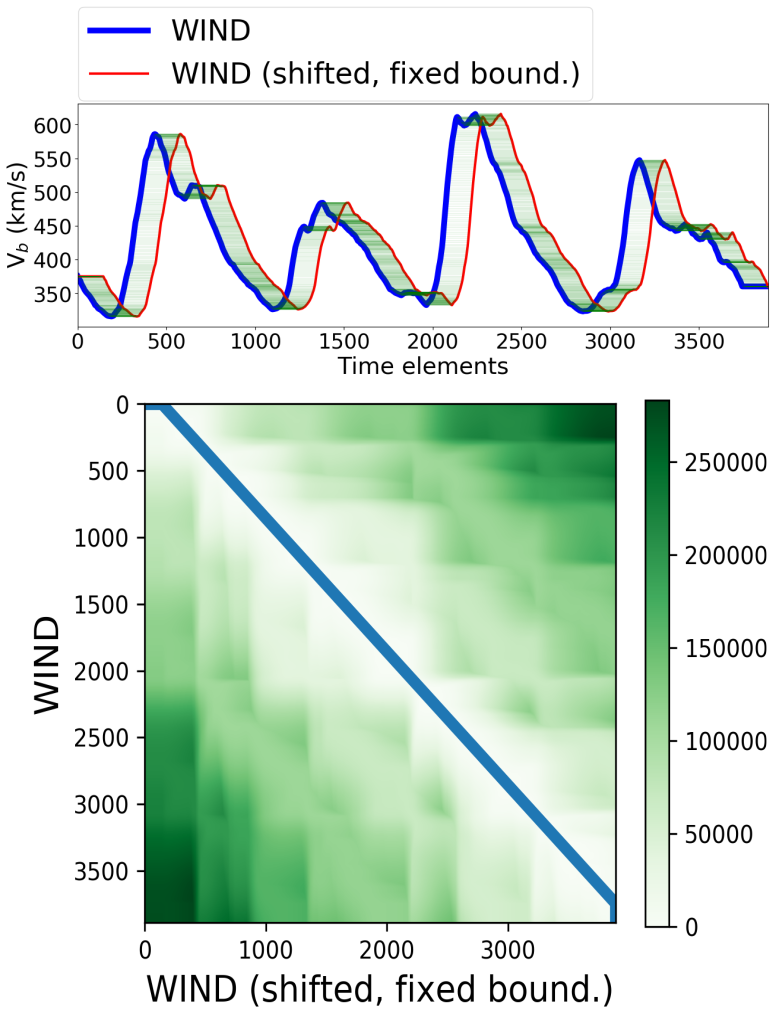}{0.3\textwidth}{(b) DTW score = 0, \\ Sum(diag) $\approx$ 18.3$\cdot$ $10^{6}$}
            \fig{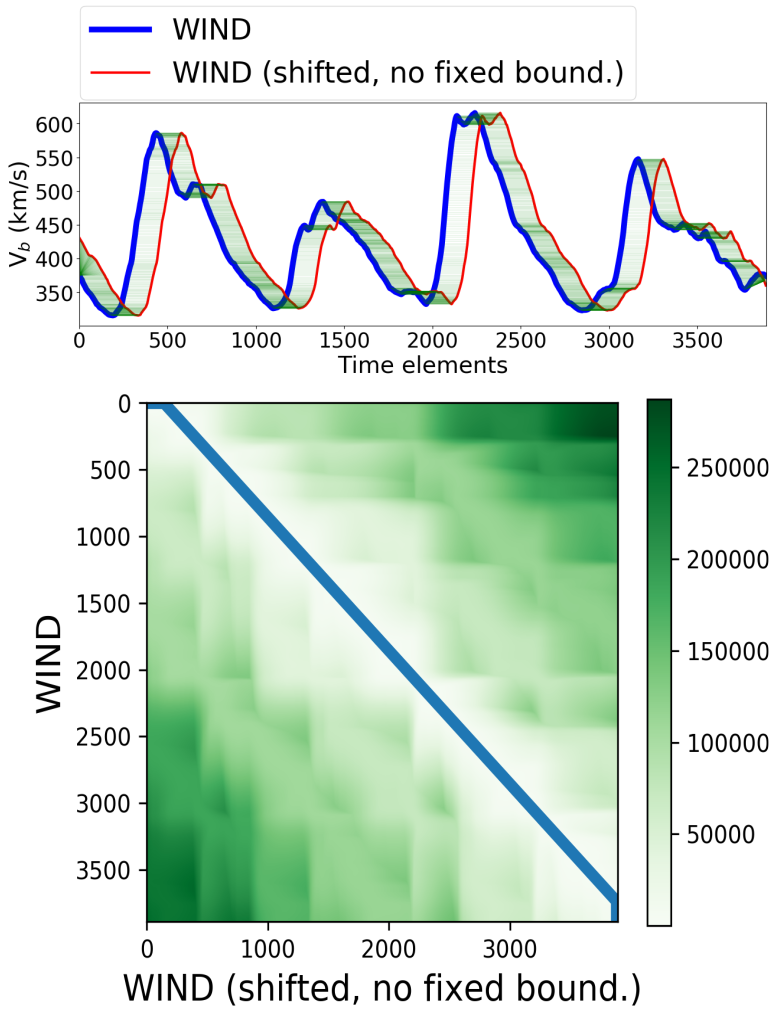}{0.3\textwidth}{(c) DTW score = 5242.86, \\ Sum(diag) $\approx$ 33.7$\cdot$ $10^{6}$}}
            
\caption{Same as Fig.~\ref{Fig:TestCases_OtherWay_all} but for time series observed by WIND during CR 2198. Panel (a): the two sequences are identical. Panel (b): the red sequence was shifted one day compared to the blue one. The initial and final points of the sequences overlap. Panel (c): the red sequence was shifted one day compared to the blue one while their initial and final points do not overlap. On the contrary, the 1-day data gap has been filled-in with data observed by WIND one day earlier. }
\label{Fig:CR2198_OtherWay_all}
\end{figure}

In Fig.~\ref{Fig:CR2198_OtherWay_all} we follow the same procedure as in Fig.~\ref{Fig:TestCases_OtherWay_all} but for sequences coming from our data-set. We employ WIND data from the time period 2017-12-03 to 2017-12-30 (CR 2198). In Fig.~\ref{Fig:CR2198_OtherWay_all}a we show how DTW behaves when comparing the observed solar wind bulk speed with the same data-set, assuming the ideal scenario in which the observed and predicted time series are exactly the same. The $x$-axis in the time series plots corresponds to the index number of the elements which actually describe time, thus the ``time elements" label. In Fig.~\ref{Fig:CR2198_OtherWay_all}b the observed data-set has been shifted forward by one day. The gap that is created between element 0 and 144 (1 day) is then filled with the value at time 0 of the non-shifted (blue) time series. The same happens for the gap that is created between element 3744 and 3888 (last day) of the non-shifted (blue) time series. It is filled with the value at time 3888 of the shifted (red) time series. The DTW score in this case is zero and the warping path follows the same pattern as described in Fig.~\ref{Fig:TestCases_OtherWay_all}b. Nevertheless, for assessing the performance between observed and predicted time series we rarely have the initial and final points of the sequences matched the way shown in Fig.~\ref{Fig:CR2198_OtherWay_all}b. The same sequences as in Fig.~\ref{Fig:CR2198_OtherWay_all}a,b are also shown in Fig.~\ref{Fig:CR2198_OtherWay_all}c, but now the shifting has occurred by filling in the one-day-shifted interval with data recorded by WIND one day earlier. As a result, the initial and final points of the sequences are not the same anymore. The DTW score in this case is 5242.86. This number does not have a meaning yet, it only reflects the fact that the cost of aligning the two time series is much bigger than in the cases of Fig.~\ref{Fig:CR2198_OtherWay_all}a,b and arises due to the alignments of the first 144 and final 144 points.

\subsection{The importance of the applied smoothing}
\label{subsection:smoothing}

\begin{figure}
\centering
\gridline{\fig{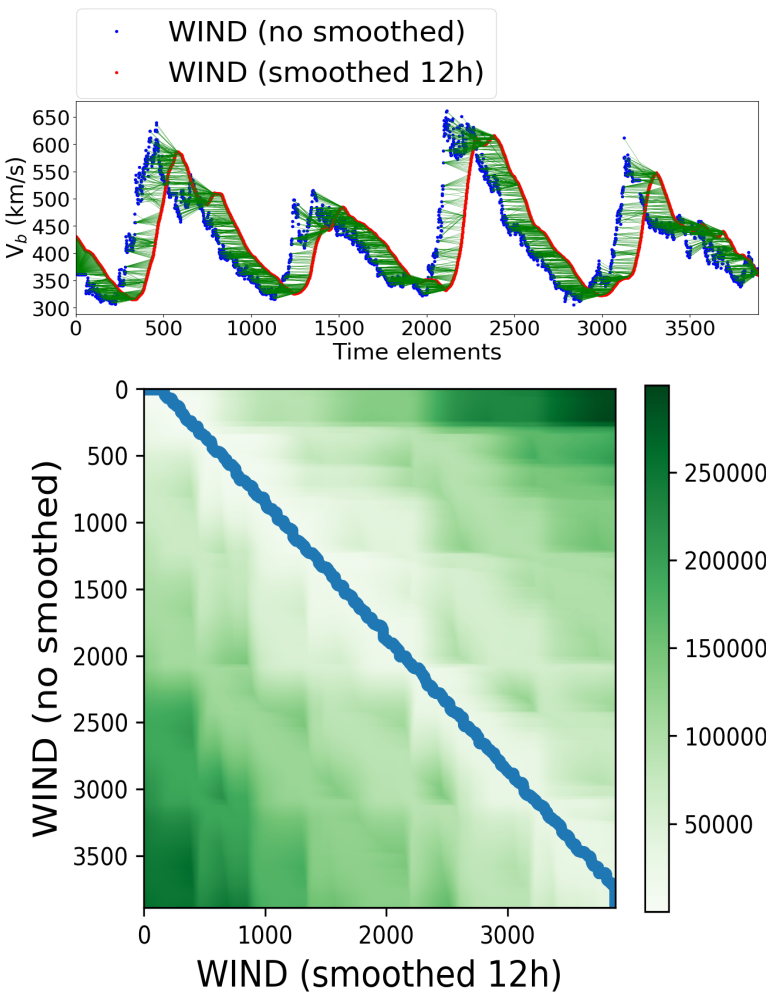}{0.4\textwidth}{(a) DTW score = 34785.24, \\ Sum(diag) $\approx$ 92.7$\cdot$ $10^{6}$}
            \fig{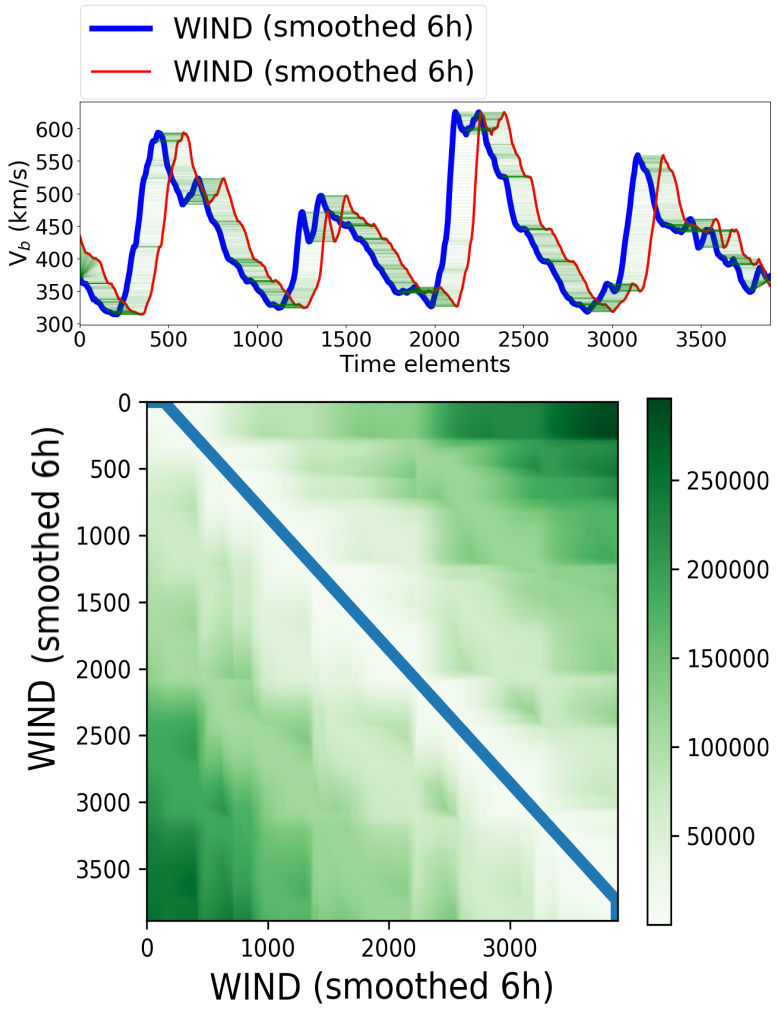}{0.4\textwidth}{(b) DTW score =  6342.19, \\ Sum(diag) $\approx$ 36.6$\cdot$ $10^{6}$}}
            
\gridline{\fig{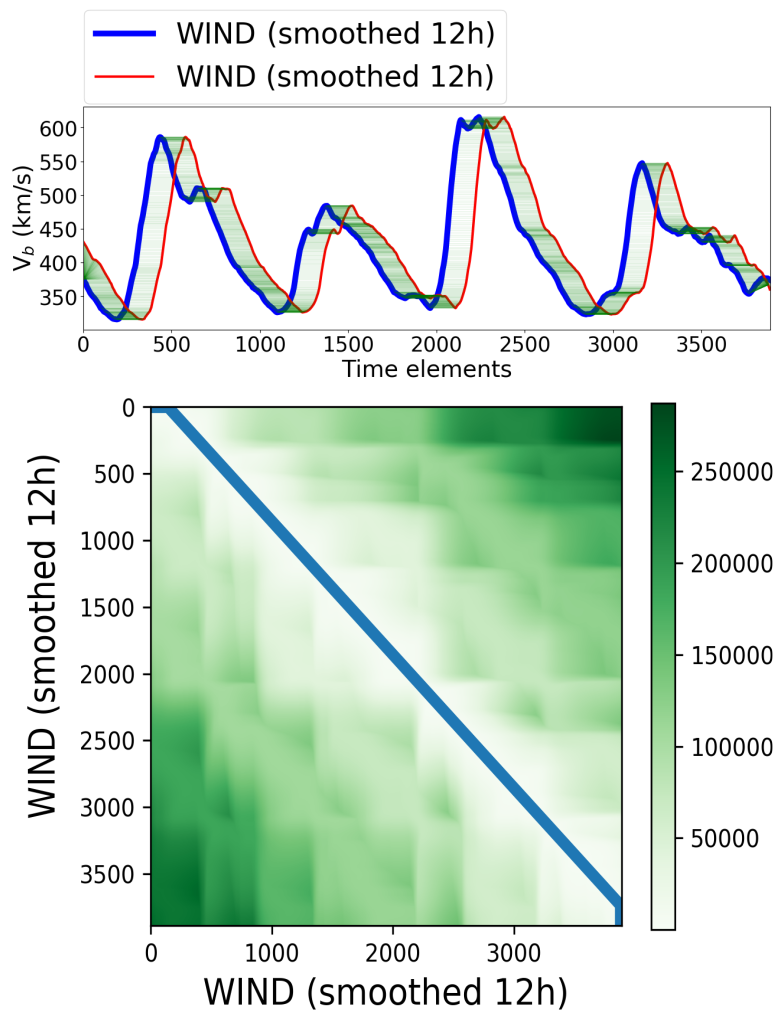}{0.4\textwidth}{(c) DTW score = 5242.86, \\ Sum(diag) $\approx$ 33.7$\cdot$ $10^{6}$}
            \fig{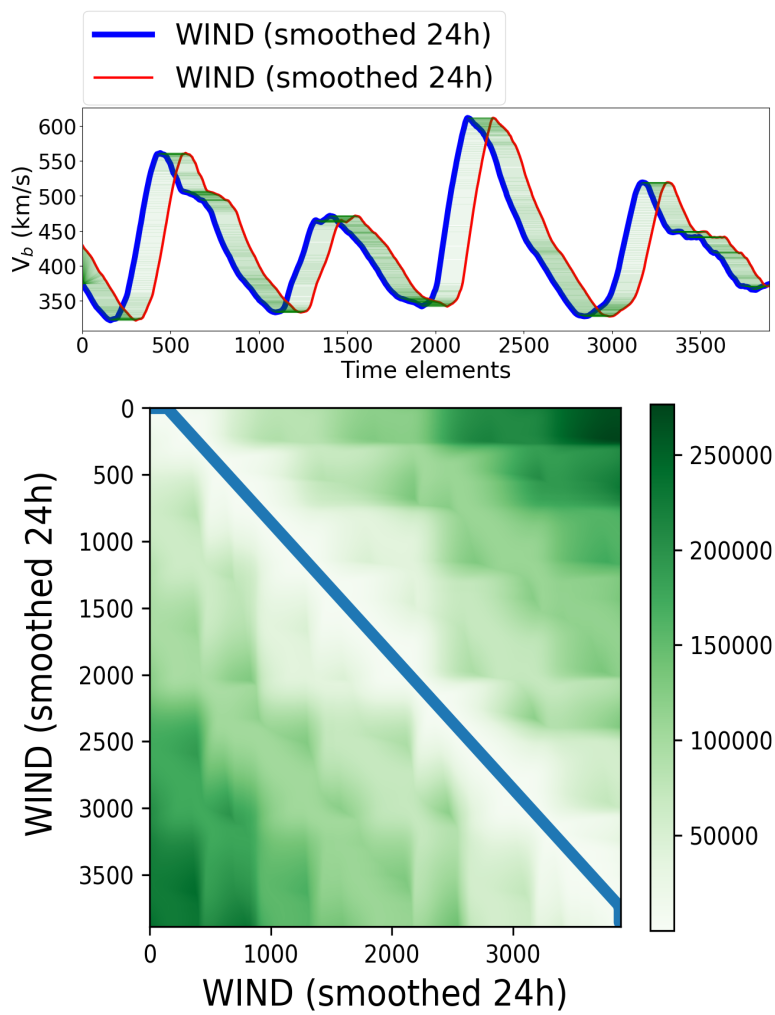}{0.4\textwidth}{(d)DTW score = 4473.49, \\ Sum(diag) $\approx$ 32.6$\cdot$ $10^{6}$}}
            
\caption{Same as Fig.~\ref{Fig:CR2198_OtherWay_all} but for time series of different smoothing. Panel (a): a comparison between time series with no smoothing and 12h smoothing is presented. Panels (b), (c) and (d): a 6h, 12h and 24h smoothing has been applied to both time series, respectively. The red sequence is always shifted 1 day compared to the blue one, similar to Fig.~\ref{Fig:CR2198_OtherWay_all}c.}
\label{Fig:DifferentSmoothings}
\end{figure}

The second important task is to apply an optimal smoothing to the observed time series. This is a subjective procedure which depends on the goals of the study, the data-set and the user himself. Usually the real data, as recorded by WIND at L1, contain high frequency fluctuations opposite to the modeled time series which are described by a smooth trend (Fig.~\ref{Fig:ExamplesDTW}). For a proper comparison between observations and predictions, it is optimal to smooth observed time series similarly to the modeled ones, as local minima and maxima influence the DTW result and generate more singularities. This happens because the method aims to match local fluctuations to parts from the modeled time series where no fluctuations have been detected. Figure~\ref{Fig:DifferentSmoothings}a shows such an example. The alignment of the points is not optimal and the cumulative DTW cost is larger than in the case where fast fluctuations have been smoothed. However, we should also be careful not to smooth out important features. An analytic example of how different smoothing influences the results is presented in Fig.~\ref{Fig:DifferentSmoothings}b, c, d. In this figure, we apply a time-centered smoothing based on a moving average of 6h, 12h and 24h to the WIND time series shown in Fig.~\ref{Fig:CR2198_OtherWay_all}c. In case the smoothing window exceeded the size of the array, we applied data padding by reflecting about the edges of the first/last elements.

Besides the different smoothing, we keep the same temporal resolution for both time series under comparison. That means that both sequences have the same amount of elements (even though DTW can also be applied to unequal time series \cite[see, e.g.,][and references therein]{Wong2003,Zhang2021}. In our analysis we adopted a 10-minute resolution and a 12-hour smoothing window as the ideal set-up for WIND and EUHFORIA time series (see section \ref{section:ApplicationsInEUHFORIA}).

\subsection{Window constraint}
\label{subsection:window_constraint}

After the time series pre-processing, we discuss the employed DTW constraints. The first restriction comes from the fact that the time warping of the sequences needs to be done within a specific time interval. We have to restrict all possible matches of the points within a specific time window, otherwise the temporal alignment between them could be indefinite. For solar wind forecasting purposes, it is undesirable for a point at day one to be matched with a point at day five, six, seven or more if the temporal uncertainty of predictions lies within a smaller time window. The approximate maximum $\Delta$t in EUHFORIA between the arrival of the observed and modeled HSSs for the time interval of interest (November 2017 - September 2019) is $\approx$ 2 days, so a time window of $\pm$2 days is applied for our purposes. Some other studies have shown that there is mismatch by at least 1 day and up to 3 days between the arrival of modeled and observed solar wind structures \citep[see e.g.,][]{owens08, macneice09, jian15, gressl14, reiss16, TemmerHintReiss2018}. Setting a window like that, also reduces the computational time in calculating the cost matrix. No constraints regarding slope weighting or step pattern are further imposed because we do not want to bias the alignment of the points towards one or another direction. 

\begin{figure}
\centering
\gridline{\fig{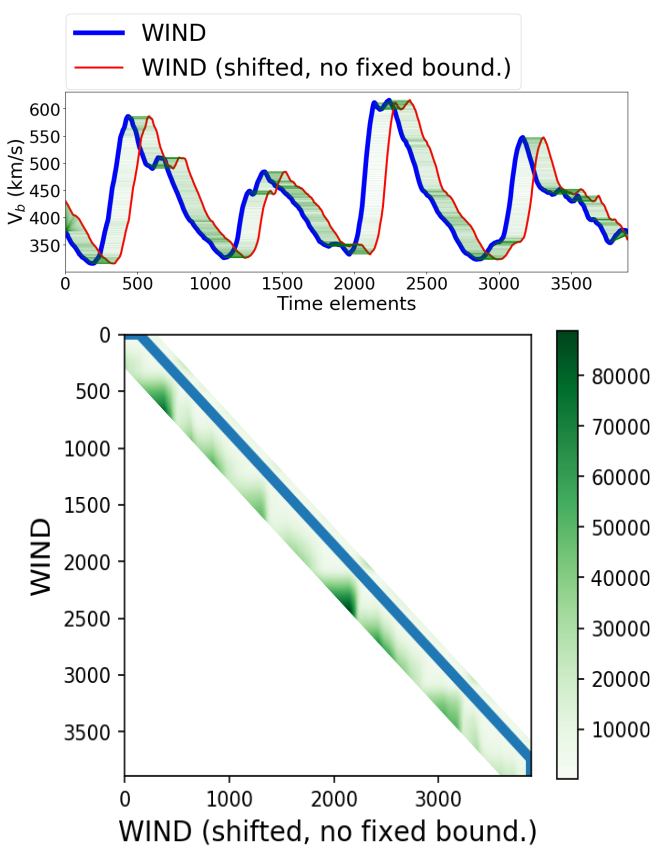}{0.4\textwidth}{(a) DTW score = 5242.86, \\ Sum(diag) $\approx$ 33.7$\cdot$ $10^{6}$}}

\caption{Same as Fig.~\ref{Fig:CR2198_OtherWay_all}c and Fig.~\ref{Fig:DifferentSmoothings}c but with a time window constraint of $\pm$2 days. }
\label{Fig:WindowConstraint}
\end{figure}

In Fig.~\ref{Fig:WindowConstraint} we show the DTW cost matrix (heat-map) and the alignment of the time series presented in Fig.~\ref{Fig:CR2198_OtherWay_all}c and Fig.~\ref{Fig:DifferentSmoothings}c, when we apply a time window constraint of $\pm$2 days. The DTW score and alignment of the points are the same, so is the warping path. The only thing that changes is the extent of the DTW matrix. We notice that it is now only filled-in within a zone along the diagonal. This is because of the 2-day window restriction we imposed that does not allow the alignment of points outside that time window. The subset of matrix that the warping path is allowed to visit is called a ``warping window" or ``band" \citep{Ratanamahatana&Keogh}. In our case, we implement a band similar to the Sakoe-Chiba band \citep{SakoeCHiba78}. The reason that the DTW score and warping path are the same in this particular example is because the maximum time difference between the two sequences is 1 day. Therefore, our results are not affected by the window constraint.

\subsection{Influence of CMEs}
\label{Influence of CMEs}

During the considered time intervals, fifteen CMEs were recorded influencing Earth. Five of them occurred very close to each other, between 2019-05-06 and 2019-05-30 \citep[see][]{RClist2003, RClist2010}. Even though most of the periods we evaluated did not include any CME, for the cases that CMEs were detected, we ignored them as events and applied DTW normally, as if these structures were not there. As a result, our recommendation for a future user who wants to assess an improved version of EUHFORIA for the same time intervals, is to work the same way as we do in this paper; namely, ignore the potential CME structures and apply DTW as if these structures never occurred. Only then, the decrease (or increase) of the DTW score will be consistent and comparable to the one calculated based on the current EUHFORIA version, so that we will be able to track how the change of the model influences the modeling output.

\section{Application of DTW for assessing the performance of solar wind time series}
\label{section:DTW_two_ways_of_evaluation}

DTW can be applied in two ways for the evaluation of modeled solar wind time series. The first way compares the DTW score of the predicted time series to an ideal and a non-ideal (reference) case scenario. The second way quantifies the time and velocity differences between each point of the time series, as aligned by DTW. 

\subsection{First way of applying DTW: the Sequence Similarity Factor (SSF)}
\label{subsection:SSF}

The WIND data for CR 2198 (blue time series in Fig.~\ref{Fig:WindowConstraint}) will be considered our observations while the same data-set, shifted by 1 day (red time series), will be our predictions. The ideal scenario will be the flawless forecast in which the red time series is identical to the blue ones (see Fig.~\ref{Fig:CR2198_OtherWay_all}a, for which DTW score = 0 and sum(diag) = 0). If the case we study is not the ideal one, then the DTW score is other than zero and does not have an actual meaning unless it is compared to (a) the ideal scenario and (b) a non-ideal (reference) prediction. We define this reference prediction as the mean model of observations which represents the forecast of the average observed speed for the period of interest. Such model has no variations with time, and will be proved useful later for the direct comparison of DTW results with traditional metrics (see Section \ref{section:ApplicationsInEUHFORIA}). 

\begin{figure}
\centering
\gridline{\fig{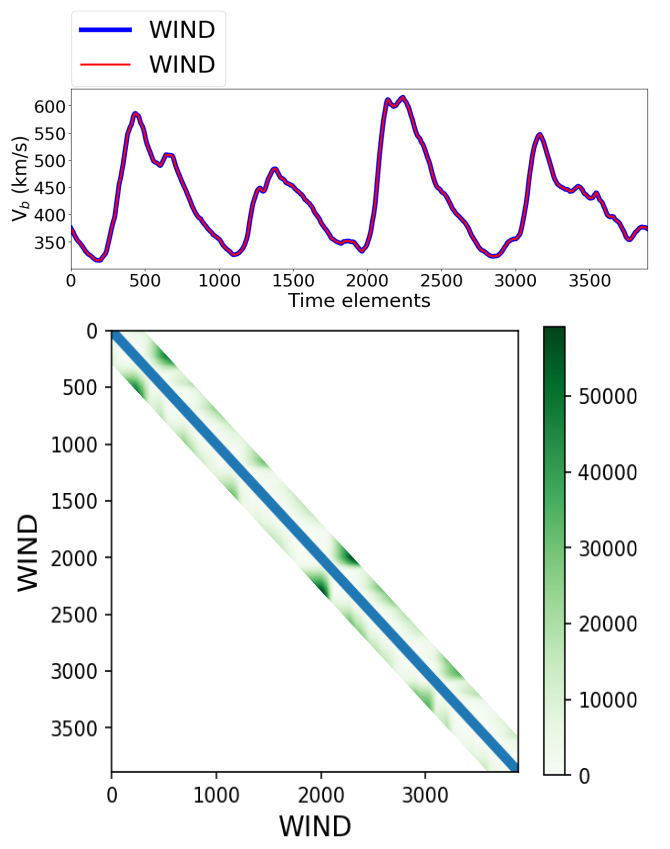}{0.3\textwidth}{(a) DTW$_{\textrm{score}}(O,M)$ = 0}
            \fig{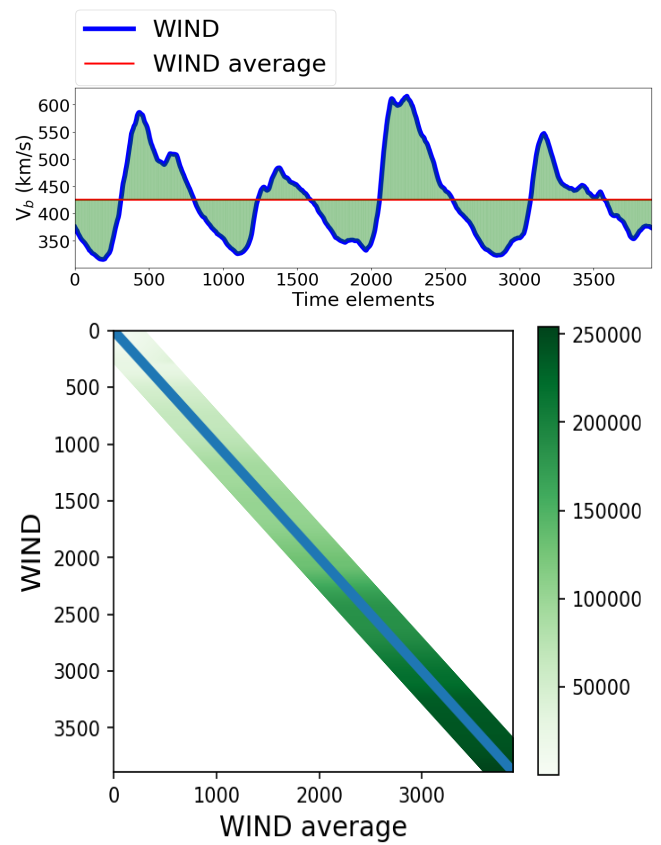}{0.3\textwidth}{(b) DTW$_{\textrm{score}}(O,\bar{O})$ = 253850.04}
          \fig{DTWplots/CR2198_window_constraint.png}{0.3\textwidth}{(c) DTW$_{\textrm{score}}(O,M)$ = 5242.86 }}
           
\caption{Example of the first approach to evaluate the performance of the predicted solar wind time series compared to observations. Panel (a): The ideal prediction scenario with DTW$_{\textrm{score}}(O,M)$ = 0. Panel (b): the non-ideal/reference case prediction scenario with the maximum DTW score for that specific time interval for which DTW$_{\textrm{score}}(O,\bar{O})$ = 253850.04. Panel (c): our actual observed and predicted time series with a DTW$_{\textrm{score}}(O,M)$ ranging between 0 and 253850.04.}
\label{Fig:SSF_plots}
\end{figure}

In Fig.~\ref{Fig:SSF_plots}a, b, c we show the application of DTW between observations and (a) the ideal prediction scenario, (b) the mean model, and (c) our predicted data-set, respectively. After calculating the DTW scores for each of these cases, we quantify the similarity of the observed and predicted time series. This is done through the Sequence Similarity Factor (SSF) which we define as: 

\begin{equation}
\centering
    \textrm{SSF} = \frac{\textrm{DTW}_{\textrm{score}}(O,M)}{\textrm{DTW}_{\textrm{score}}(O,\bar{O})}, \quad \textrm{SSF}\in [0,\infty].
\label{eq:SSF}
\end{equation}

\noindent where $O$, $M$ and $\bar{O}$ stand for $``Observed"$, $``Modeled"$ and $``Averaged$ $Observed"$ data, respectively. The SSF is equal to zero when we have achieved the perfect forecast, and equal to one, when our forecast is as bad as a straight, average line prediction. In Fig.~\ref{Fig:SSF_plots}c, the SSF between the observed and predicted time series is 0.021, very close to the perfect scenario of SSF = 0 (Fig.~\ref{Fig:SSF_plots}a).

\begin{figure}
\centering
\gridline{\fig{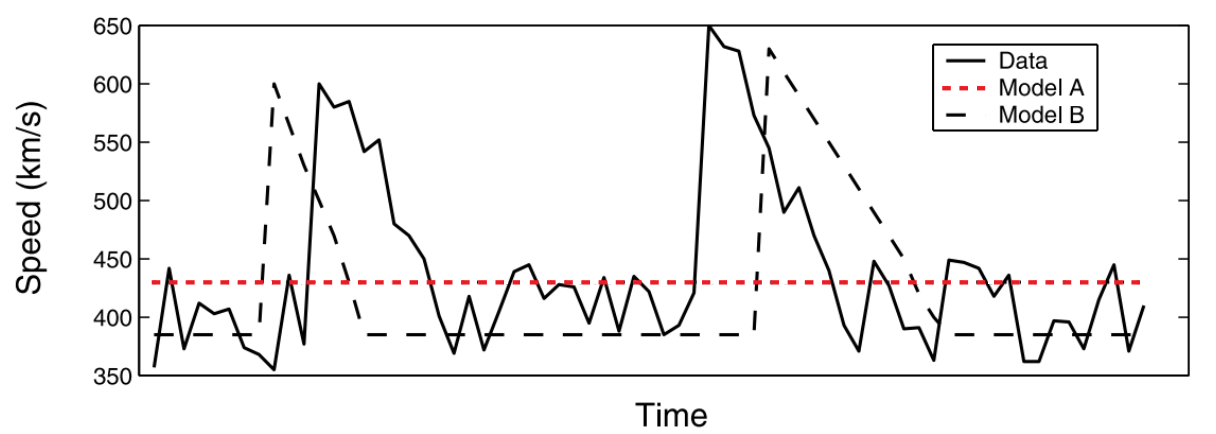}{0.8\textwidth}{}}
           
\caption{Example of an observed data-set (solid line) and two predicted time series (Model A, red-dashed line and Model B, black-dashed line). Model A is just a straight line while Model B is very similar to the observed data but shifted in time. The MSE of Model B is larger than the one calculated for Model A, meaning that a straight-line prediction performs better than a prediction which is very similar to observations, but shifted in time \cite[adapted from][]{Owens05}.}
\label{Fig:OwensFigure}
\end{figure}

The fact that DTW warps dynamically the sequences in time and, as a result, it can locate which point from one time series better corresponds to a point from the other time series is a huge advantage compared to other metrics. In their study, \citet{Owens05} noted that one of the most often used metric, the mean square error (MSE), has a very significant drawback even if it is a useful tool for a first-order assessment of time series. This drawback comes from the fact that a straight line (Model A, red-dotted line in Fig.~\ref{Fig:OwensFigure}) can sometimes give a lower MSE when compared to observations, from a time series that is very similar to observations but just shifted in time (Model B, black-dashed line in Fig.~\ref{Fig:OwensFigure}). DTW overcomes this problem, opposite to the simple error functions which are completely based on the Euclidean distance measure (see, e.g., the comparison between the DTW scores calculated in Fig.~\ref{Fig:SSF_plots}b and c and the example shown in Fig.~\ref{Fig:OwensFigure}). Nevertheless, there are still some cases for which the straight average line performs better than our modeled data-set. For these cases, it is not the potential shifting in time that causes this discrepancy, as this has already been solved by DTW. On the contrary, the variability in the $y$-axis which is sometimes opposite to what is observed in the real data (e.g., the model predicts valleys in place of peaks, and vice versa), is the reason why we get DTW scores which are bigger than the ones calculated for the straight-line scenario. As a result, it is reasonable for such cases to obtain a SSF which is higher than one.


\subsection{Quantification of time and amplitude differences} 

\begin{figure}
\centering
\gridline{\fig{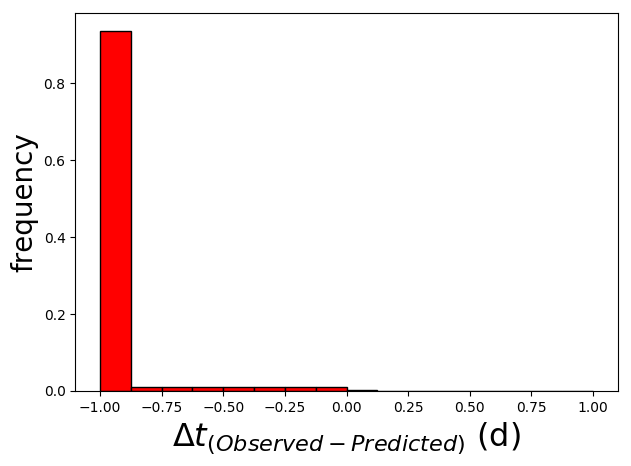}{0.4\textwidth}{(a)}
        \fig{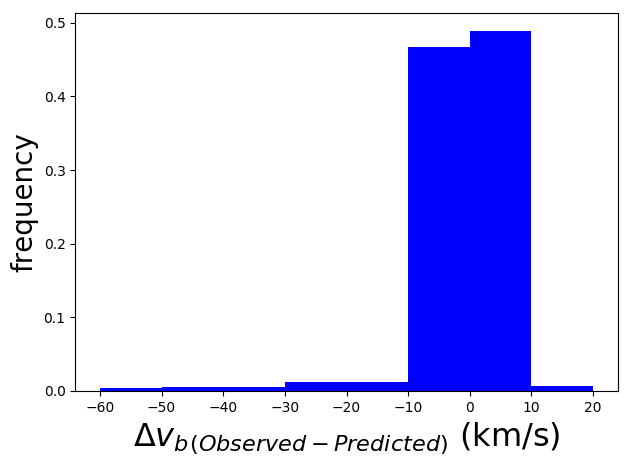}{0.4\textwidth}{(b)}}
           
\caption{Histograms of time and velocity differences between the aligned points, as matched by DTW. Panel (a): histogram of time differences ($\Delta$t in days). Panel (b): histogram of velocity differences ($\Delta$v$_{b}$ in km/s).}
\label{Fig:Histos}
\end{figure}

Besides the SSF, DTW permits the estimation of time differences and amplitude differences between the points that are best aligned. This important feature is not easily provided by other metrics which can usually quantify only one of those aspects at a time. Figure~\ref{Fig:Histos}a, b shows the histograms of time and amplitude (velocity) differences between the aligned points of the sequences presented in Fig.~\ref{Fig:SSF_plots}c. A maximum $\Delta$t of 1 day can be observed in Fig.~\ref{Fig:Histos}a, which was expected since this is how much we shifted our data. Figure~\ref{Fig:Histos}b shows a maximum $\Delta$v$_{b}$ of 60 km/s. The maximum difference in velocity arises at the beginning of the time series when our predicted data-set (red time series) is higher than the observed one (blue time series).

We note that the alignments of the points provided by DTW do contain singularities, i.e., when a point from one sequence can be matched with two or more points from the other sequence. Even though singularities are generally assumed as a drawback of DTW, we find their existence necessary as they help ensuring the best possible alignment between time series. As a result, the $\Delta$t and $\Delta$v$_{b}$ presented in the histograms are not only relevant to the time and amplitude differences of the points that are aligned uniquely between them, but to the singularity points, as well. By improving the model and consequently, the solar wind forecasting, these singularities are minimized, the DTW score becomes lower and a better agreement with the observations is achieved. 

\section{Evaluating the performance of solar wind time series in EUHFORIA}
\label{section:ApplicationsInEUHFORIA}
\subsection{SSF versus traditional skill scores: using the mean model as a reference model}
\label{section:DTWvsMSE_meanmodel}

\begin{figure}
\centering
\gridline{\fig{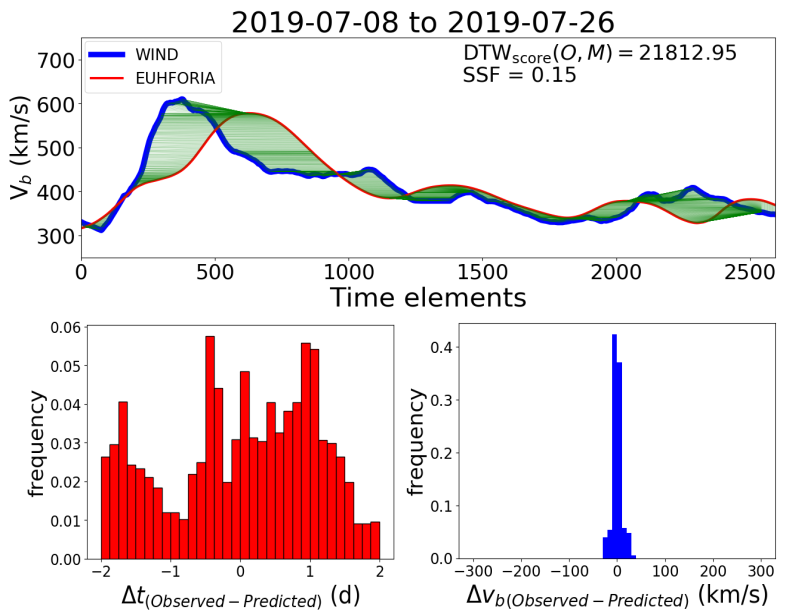}{0.8\textwidth}{}}
           
\caption{DTW alignments and histograms of time and velocity differences between 2019-07-08 to 2019-07-26 (period 20) during which EUHFORIA performed the best compared to all other considered periods. The time in the x-axis of the upper panel corresponds to evenly spaced time elements with a time difference of 10-min between each other.}
\label{Fig:CarRot23_best}
\end{figure}

In this section, we apply the DTW method on the solar wind velocity time series as modeled by EUHFORIA v1.0.4, for the period November 2017 to September 2019. The set-up of EUHFORIA is the same as the one presented in \citep{pomoell18, Hinterreiter19}. We first split the considered time interval in smaller periods listed in the second column of Table \ref{Table:DTWTable}. Then, we adopt smoothing and window constraints (see Sections \ref{subsection:smoothing} and \ref{subsection:window_constraint}, respectively). Finally, we quantify EUHFORIA's performance compared to observations by employing both DTW ways mentioned in section 4. The upper panel of Fig.~\ref{Fig:CarRot23_best} shows the DTW alignment for the time period with the lowest SSF. This corresponds to the time interval between 2019-07-08 to 2019-07-26 (period 20) during which EUHFORIA has performed the best compared to all other periods we considered in this study. In the same figure, we also present the histograms of time and velocity differences between observed and predicted data-sets. The time difference between the two sequences is $\pm$2 days, which is the maximum temporal window we imposed for the alignment. The velocity differences are the lowest among all other periods, with a maximum of 40 km/s. The application of DTW to the rest of the periods are shown in Appendix \ref{appendix:DTWalignments&histos} and the SSFs are summarized in the fourth column of Table \ref{Table:DTWTable}.


To prove our point about the significant advantages that DTW offers compared to traditional metrics (see section \ref{section:DTW_two_ways_of_evaluation}), in the fifth column of Table \ref{Table:DTWTable} we present the results of EUHFORIA's performance compared to observations as evaluated based on a traditional MSE-based skill score metric. This metric is defined as: 

\begin{equation}
\centering
    \textrm{Skill} \: \textrm{Score} = \frac{\textrm{MSE}}{\textrm{MSE}_{\textrm{ref}}}, \quad \textrm{Skill} \: \textrm{Score}\in [0,\infty].
\label{eq:SkillScore}
\end{equation}

\noindent The nominator corresponds to the MSE between observations and EUHFORIA while the denominator corresponds to the MSE between observations and a reference model. For this section, the reference model will be the mean model. Similarly to eq. \ref{eq:SSF}, a Skill Score equal to zero corresponds to the perfect prediction while a Skill Score equal to one means that the prediction performs the same as the reference model. For a Skill Score higher than one, the reference model performs better than our predictions.

\begin{figure}
\centering
\gridline{\fig{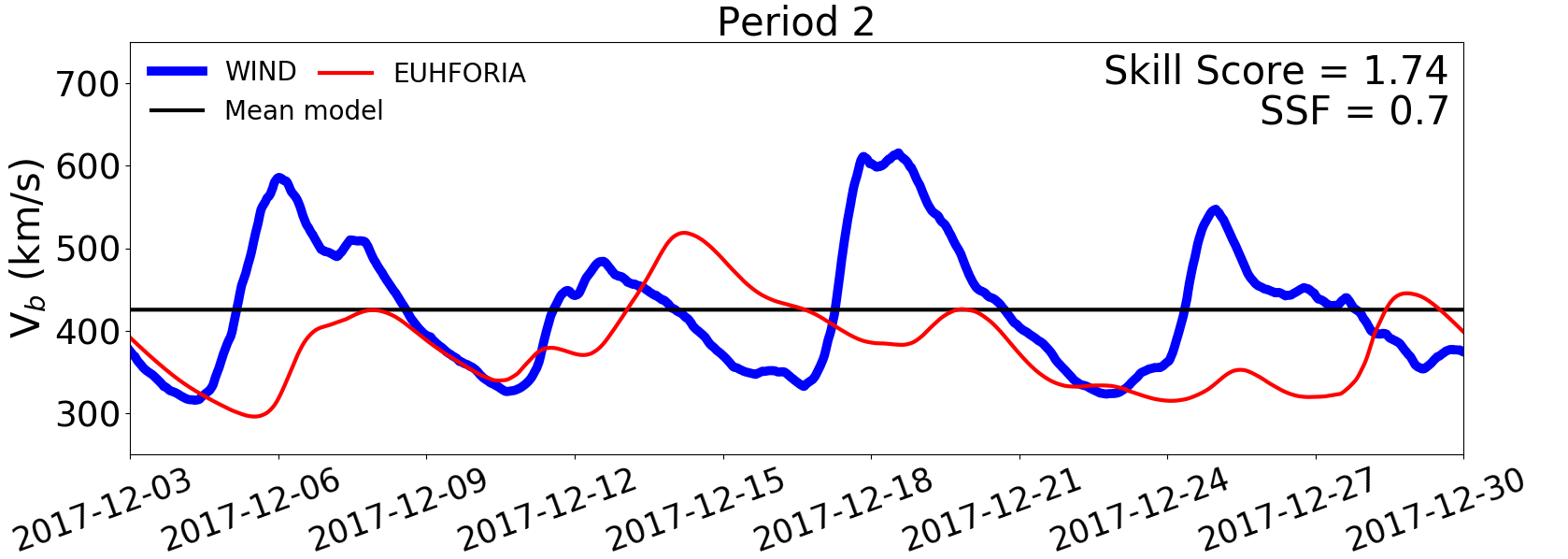}{0.49\textwidth}{(a)}
        \fig{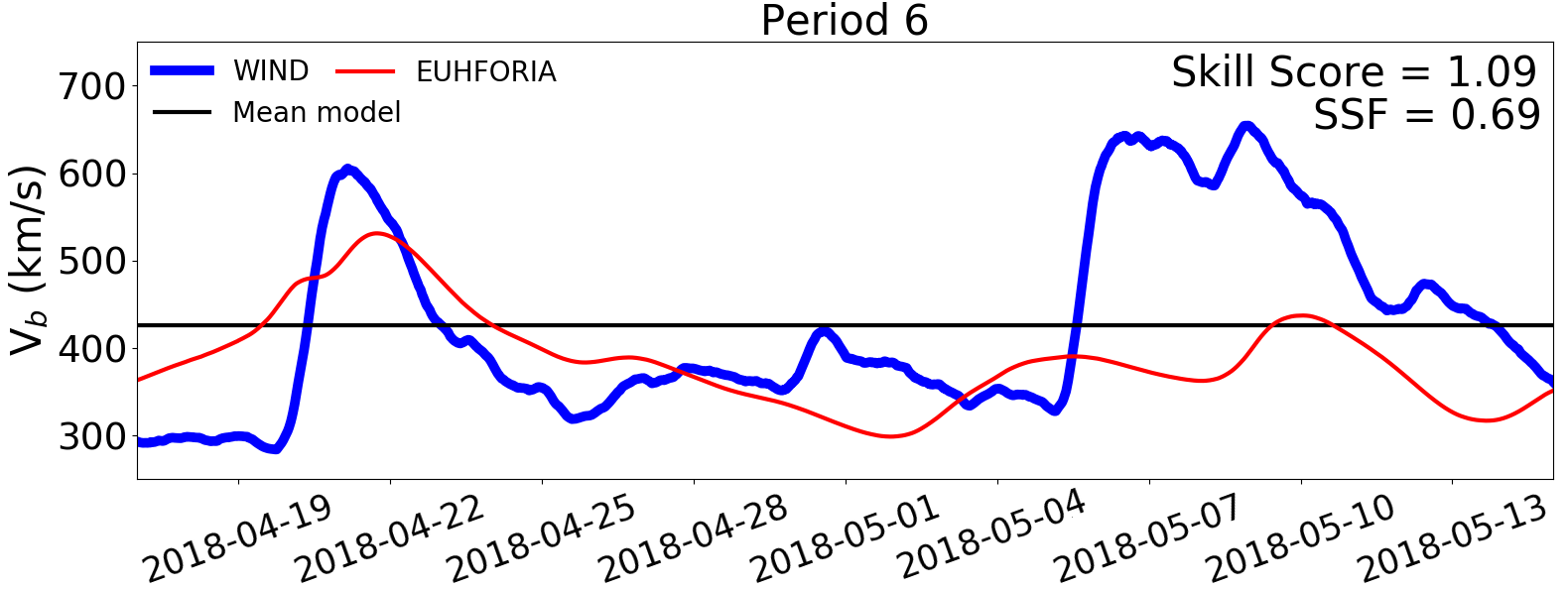}{0.47\textwidth}{(b)}}
           
\caption{WIND observations (blue), EUHFORIA output (red) and mean model (black) are shown for periods 2 (panel a) and 6 (panel b). The SSF and traditional Skill Score are also presented in the upper right part of the figures.}
\label{Fig:DTWvsMSE_MeanModel}
\end{figure}

Comparing the SSF and Skill Score values in columns four and five of Table \ref{Table:DTWTable}, we see that even though both measures have been defined similarly, they provide different assessment results for the same periods. Periods 2 and 6 correspond to such controversial examples for which SSF $<$ 1, meaning that EUHFORIA performed better than the mean model, but at the same time Skill Score $ >$ 1, meaning that EUHFORIA performed worse than the mean model. To clarify the situation, we present the time series for these particular periods in Fig.~\ref{Fig:DTWvsMSE_MeanModel}. We notice that in both cases EUHFORIA predicted much better the observations compared to the mean model. As a result, we should expect a SSF $<$ 1 and a Skill Score $<$ 1. Nevertheless, due to the time lag of the predicted time series compared to WIND data and the inability of the traditional skill score to capture the overall shape of the sequence, the MSE of the mean model was lower than the MSE of the EUHFORIA time series. This resulted to a Skill Score $ >$ 1 which does not reflect the actual bad performance of the mean model in terms of forecasting the variability in the solar wind.

\subsection{SSF versus traditional skill scores: using the 27-days persistence model as a reference model}

In this section, we perform the same test as in section \ref{section:DTWvsMSE_meanmodel} but we employ a different reference model. That is, the 27-days persistence model \citep[][]{Owens2013} which is widely used within the space weather community. Based on this model, we assume that the solar wind speeds measured over a full solar rotation predict the future solar rotation, as well. The SSF and Skill Score results for the individual periods of interest, are summarized in the sixth and seventh columns of Table \ref{Table:DTWTable}. We notice that in all but three periods (period 3, 8 and 10) the SSF and Skill Score agree with the evaluation of EUHFORIA time series. Namely, when SSF $<$ 1 and Skill Score $<$ 1, both measures reflect that EUHFORIA performs better compared to the 27-days persistence model, and vice versa.

\begin{figure}
\centering
\gridline{\fig{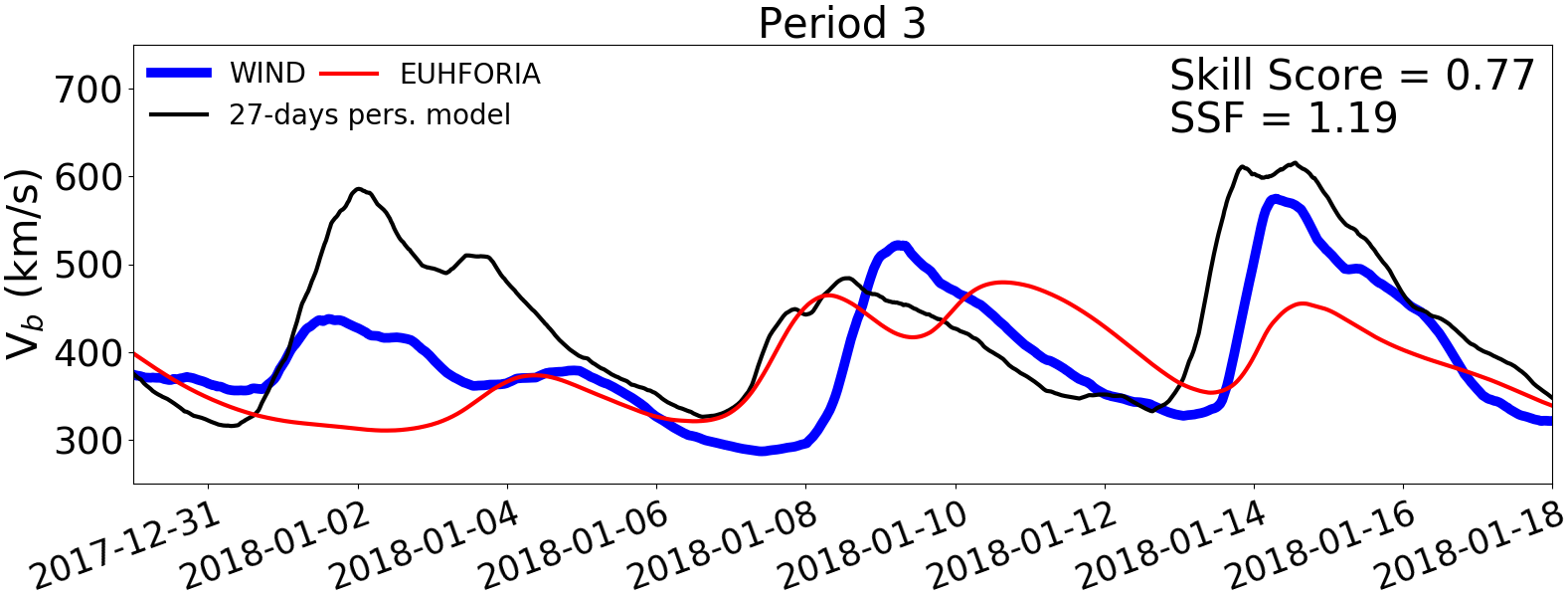}{0.49\textwidth}{(a)}
        \fig{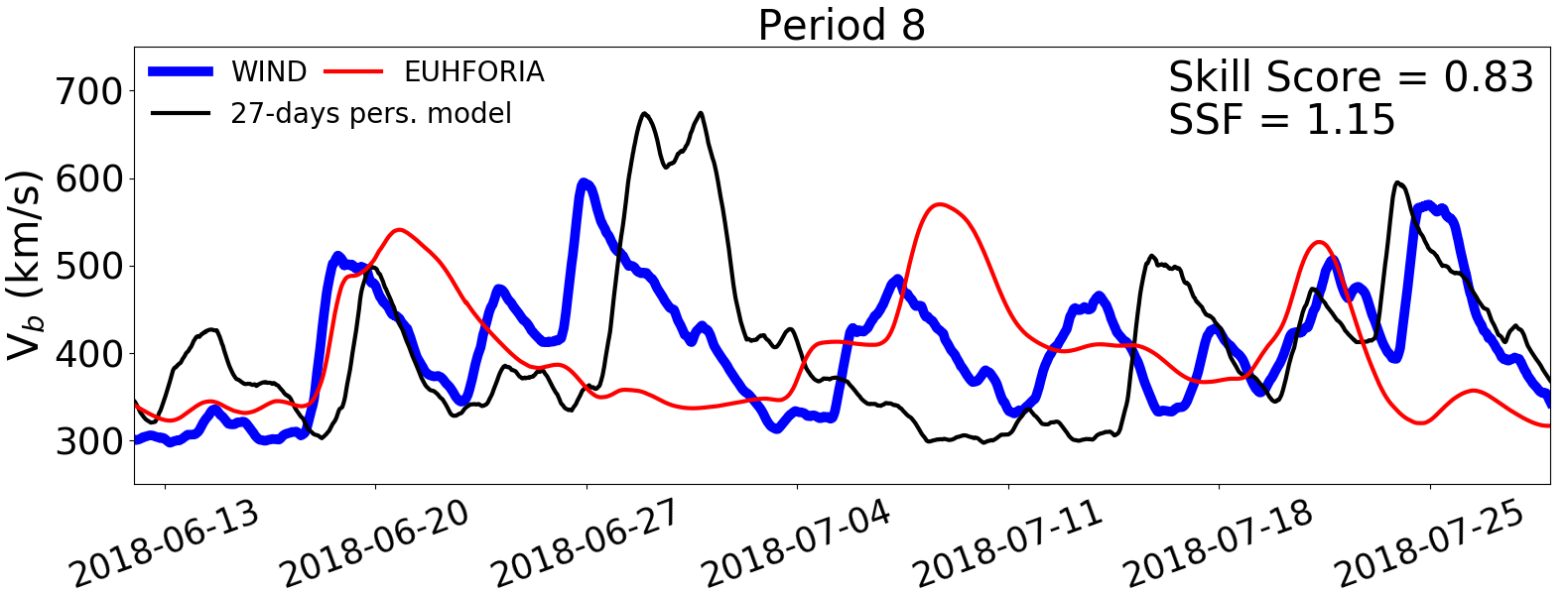}{0.5\textwidth}{(b)}}
\gridline{\fig{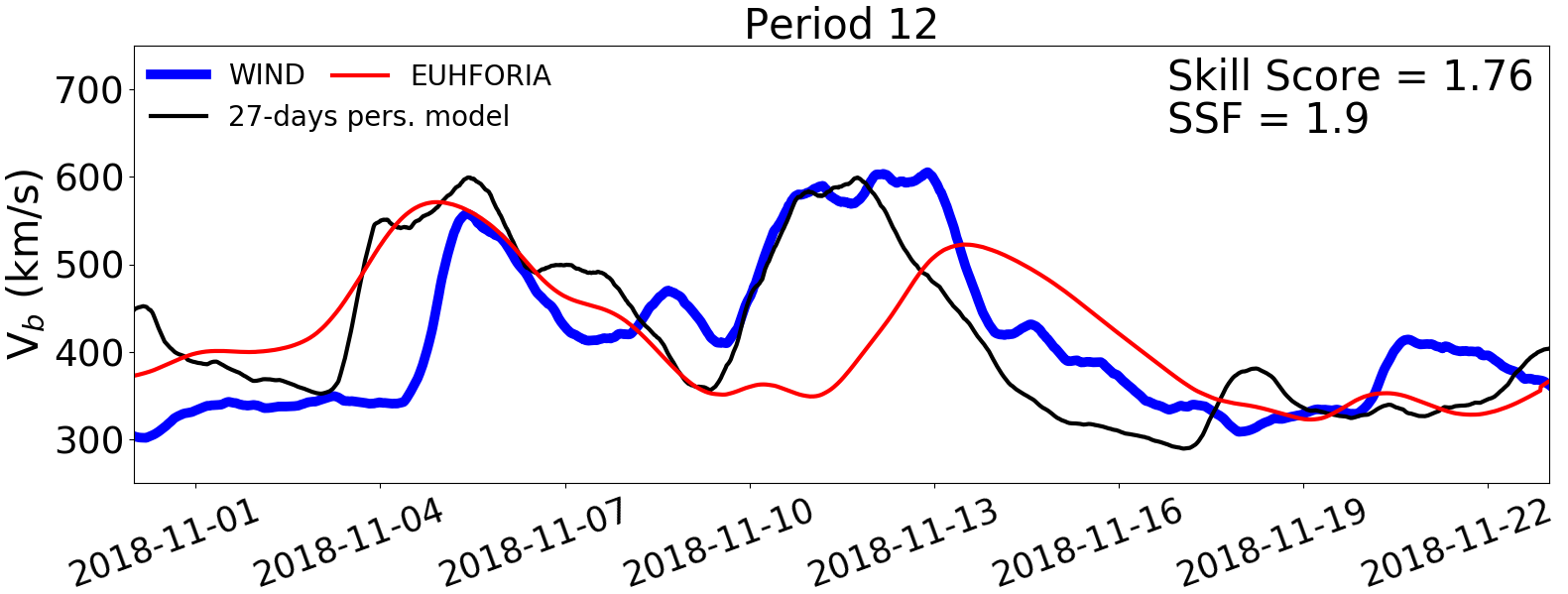}{0.49\textwidth}{(c)}
        \fig{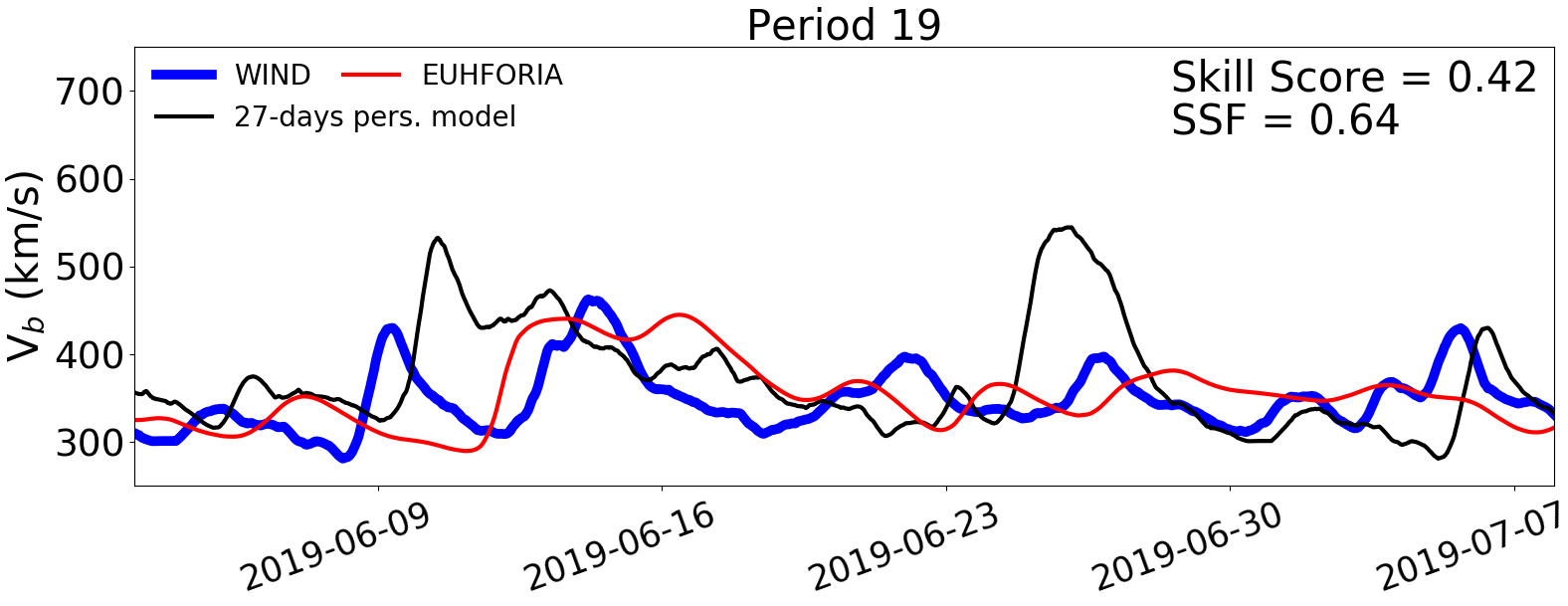}{0.49\textwidth}{(d)}}
           
\caption{WIND observations (blue), EUHFORIA output (red) and 27-days persistence model (black) are shown for period 3 (panel a), 8 (panel b), 12 (panel c) and 19 (panel d). The SSF and Skill Score are also presented in the upper right part of the figures.}
\label{Fig:DTWvsMSE_27days}
\end{figure}

In Fig.~\ref{Fig:DTWvsMSE_27days}a,b we present the time series for two out of three periods for which SSF and Skill Score showed opposite results. In both cases the 27-days persistence model predicts the WIND data better than EUHFORIA, not only because of the number of HSSs that it captures, but also because of better reproducing the HSS amplitudes. In Fig.~\ref{Fig:DTWvsMSE_27days}c,d we further show two cases for which the SSF and Skill Score agree. In the former case, the 27-days persistence model performs better than EUHFORIA and that is reflected to the SSF and Skill Score numbers. In the latter case, EUHFORIA performs better than the persistence model, with both SSF and Skill Score being lower than unity.

\begin{table}[h!]
\centering
\renewcommand{\arraystretch}{1.5}
\begin{tabular}{|c c c| c c |c c| }
\hline \hline

Period & Dates & No. of elements & SSF$_{mean}$ & Skill Score$_{mean}$ & SSF$_{27-days}$ & Skill Score$_{27-days}$ \\ [0.5ex] 

\hline \hline 
1 & 2017-11-06 to 2017-12-03 &3889& 1.07 &  2.59&  4.33 & 4.46 \\
2 & 2017-12-03 to 2017-12-30 &3889& \textbf{0.70} & \textbf{1.74} &  2.99 &  2.00   \\
3 & 2017-12-30 to 2018-01-18 &2737&  0.49&  0.94&  \textbf{1.19} &  \textbf{0.77}  \\
4 & 2018-01-18 to 2018-03-07 &6913&  1.43&  3.91&  2.78 &  3.19  \\
5 & 2018-03-07 to 2018-04-17 &5905&  \textbf{0.83}&  \textbf{1.93}&  1.70 &  1.34 \\
6 & 2018-04-17 to 2018-05-16 &4177&  \textbf{0.69}&  \textbf{1.09}&  1.93 &   1.47  \\
7 & 2018-05-16 to 2018-06-12 &3889&  \textbf{0.66}&  \textbf{1.12}&  1.62 &   1.30  \\
8 & 2018-06-12 to 2018-07-29 &6769&  \textbf{0.80}&  \textbf{1.65}&  \textbf{1.15} &   \textbf{0.83}  \\
9 & 2018-07-29 to 2018-09-02 &5040&  \textbf{0.87}&  \textbf{1.56}&  1.68 &   1.73  \\ %
10 & 2018-09-02 to 2018-09-29 &3889&  0.42&  0.90& \textbf{1.03} &  \textbf{0.72}   \\
11 & 2018-09-29 to 2018-10-31 &4608&  0.43&  0.78& 0.66 &   0.80 \\ %
12 & 2018-10-31 to 2018-11-23 &3313&  \textbf{0.48}&  \textbf{1.15}&  1.90 &   1.76  \\ %
13 & 2018-11-23 to 2018-12-24 &4465&  0.47&  0.92&  1.37 &    1.29 \\
14 & 2018-12-24 to 2019-01-21 &4033&  1.07&  1.81&  1.89 &    1.07  \\
15 & 2019-01-21 to 2019-02-19 &4177&  \textbf{0.36}&  \textbf{1.02}&  1.92 &    2.88 \\
16 & 2019-02-19 to 2019-03-24 &4753&  \textbf{0.90}&  \textbf{1.41}&  1.69 &    1.05 \\
17 & 2019-03-24 to 2019-04-19 &3745&  1.09&  2.73&  1.90 &    1.88 \\
18 & 2019-04-19 to 2019-06-03 &6481&  1.17&  2.54&  1.75&    2.07 \\
19 & 2019-06-03 to 2019-07-08 &5041&  \textbf{0.63}&  \textbf{1.56}&  0.64&     0.42\\
20 & 2019-07-08 to 2019-07-26 &2593&  0.15&  0.64&  0.25&    0.54 \\
21 & 2019-07-26 to 2019-08-22 &3889&  0.49&  0.80&  1.52&     1.54\\
22 & 2019-08-22 to 2019-09-19 &4033&  \textbf{0.86}&  \textbf{1.06}&  2.89&    1.37 \\
\hline \hline
\end{tabular}

\caption{Evaluation of the performance of EUHFORIA solar wind time series based on the SSFs and Skill Scores for the individual periods of interest. A 10-minute resolution and a 12-hour smoothing window was adopted. The subscripts \textit{mean} and \textit{27-days} indicate which reference model was employed every time. The bold values correspond to cases at which the SSF and Skill Score provided opposite performance evaluations for the time series of interest.}
\label{Table:DTWTable}
\end{table}

\section{Conclusions and discussion}

In this study, we introduced an alternative way to assess the performance of solar wind time series, the so-called Dynamic Time Warping (DTW) technique. Although DTW is not a metric by definition, as it violates the triangular inequality, it acts like one. It obeys to the rule of continuity, monotonicity and the fact that the first and last points of one sequence should be matched at least with the first and last points from the other. It calculates a cumulative cost, the so-called DTW score, which represents the cost of aligning two time series in time when their pattern is similar but differ in time. DTW was already used in other disciplines, but it is the first time it is adapted and applied for the purposes of evaluating the performance of solar wind time series.

We discussed the benefits and restrictions of the technique, and presented two complementary ways that DTW can be exploited to assess the solar wind predictions provided by a model. The first way calculates the DTW score between observations and predictions, as well as between observations and a reference model. We define the ratio of these scores as the sequence similarity factor (SSF). This is a skill score which is equal to zero for a perfect forecast, equal to one when the forecast performs the same as the reference model, and higher than one when the model’s prediction is even worse than the prediction from the reference model. The second way that DTW can be exploited, is by evaluating time and amplitude differences between the points aligned by the method. As a result, DTW can be used as a hybrid metric between continuous measurements (such as, e.g., the correlation coefficient), and point-by-point comparisons by assessing simultaneously time and amplitude differences, a property not often met in traditional metrics.

We then assessed the performance of solar wind predictions by EUHFORIA for an interval of approximately two years (November 2017 - September 2019). This interval was first divided into smaller periods for faster and more accurate evaluation. To acknowledge the advantages of DTW and understand the differences with traditional skill score metrics, we performed two tests: first, we evaluated our predictions based on the SSF and a MSE-based skill score metric by employing the mean model as a reference model. Second, we repeated the same procedure but this time we employed the 27-days persistence model as reference. The former test showed that in 50$\%$ of the cases (11 out of 22 periods), the SSF and MSE-based skill score yielded opposite results. Particularly, the MSE-based skill score indicated that the mean model performed better than EUHFORIA simulations even though EUHFORIA was reproducing better the observations. The discrepancy between the two measures arises from the ability of DTW to dynamically warp the sequences in time and locate which point from one sequence better corresponds to a point from another sequence, opposite to conventional Euclidean metrics. Employing next the 27-day persistence model as the reference model in our study, we concluded that it performed better than EUHFORIA in predicting the observations, for 19 out of 22 periods of interest. In 3 out of those 19 cases, the SSF and MSE-based skill score resulted in opposite assessments, with the latter metric providing misleading evaluations for the predictions due to its inability to capture the overall shape of the time series. Therefore, we prove that DTW can be used as an objective quantification measure for model evaluation. In addition to MSE (and to other traditional metrics), it covers more detailed information on the similarity between the profile of two data sets, thus, it should be used in conjunction with other measures to provide the most complete picture of a model's performance.  

The use of DTW can also be extended to the evaluation of other solar wind signatures besides velocity, such as the solar wind density, temperature, magnetic field, pressure etc. Moreover, an extention of DTW, the so-called multi-dimensional DTW \citep[see][and references therein]{MultiDimensionalDTW, PDF_Keogh} permits the assessment of multiple sequences, i.e., in case of multi-dimensional simulations in which time series from different locations around Earth (or any other point of interest), are considered. The technique enables pattern comparisons between multiple time series at the same time which is particularly useful for the evaluation of spatial uncertainties during the arrival of a HSS at a measuring satellite. The multi-dimensional DTW could also be extended to the identification of HSSs by evaluating the various signatures of plasma and magnetic characteristics. For example, during the arrival of a HSS at a particular point of interest, the solar wind density, temperature, pressure, magnetic field and IMF polarity should conform to specific patterns and behaviors \citep[see][for more details]{Jian2006} that should be recognized by the method. These ideas have not been tested in the frame of the current study, but they consist promising ideas for the future.

For consistent evaluations of modeled solar wind time series with DTW, we recommend avoiding periods which include potential CME influence at Earth. If this is not possible, the user can still apply the DTW by ignoring the CME signatures. The DTW score will be slightly different then, compared to the case that there were no CMEs. As mentioned in section \ref{Influence of CMEs}, fifteen CMEs were identified influencing Earth between November 2017 -- September 2019. For most of the individual periods listed in Table \ref{Table:DTWTable}, there were no observed CMEs. However, for the cases that CMEs were detected, we ignored them for the application of DTW. Therefore, a future user who wants to assess an improved version of EUHFORIA for the same time intervals, should work the same way; namely, treat the potential CME structures as if they were not there. Only then, the decrease (or increase) of the DTW score will be consistent and comparable to the one calculated based on EUHFORIA v1.0.4, so that we will be able to track how the change of the model influences the modeling output. We note that DTW could be applied separately to CME time series, in terms of evaluation of the shock, the B$_{z}$ component or other in situ parameters.

\clearpage

\appendix

\section{DTW alignments and histograms between November 2017 - September 2019}
\label{appendix:DTWalignments&histos}

\begin{figure}[ht]
\centering
\gridline{\fig{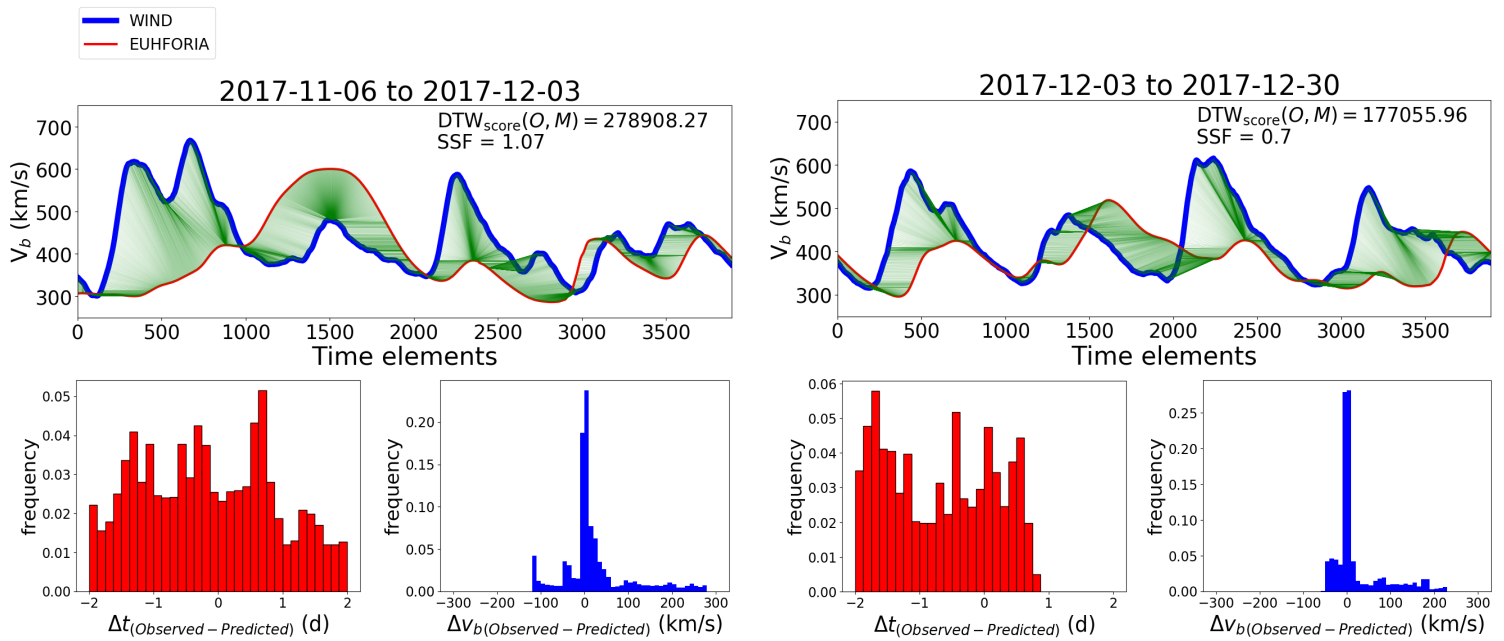}{0.95\textwidth}{}}
\gridline{\fig{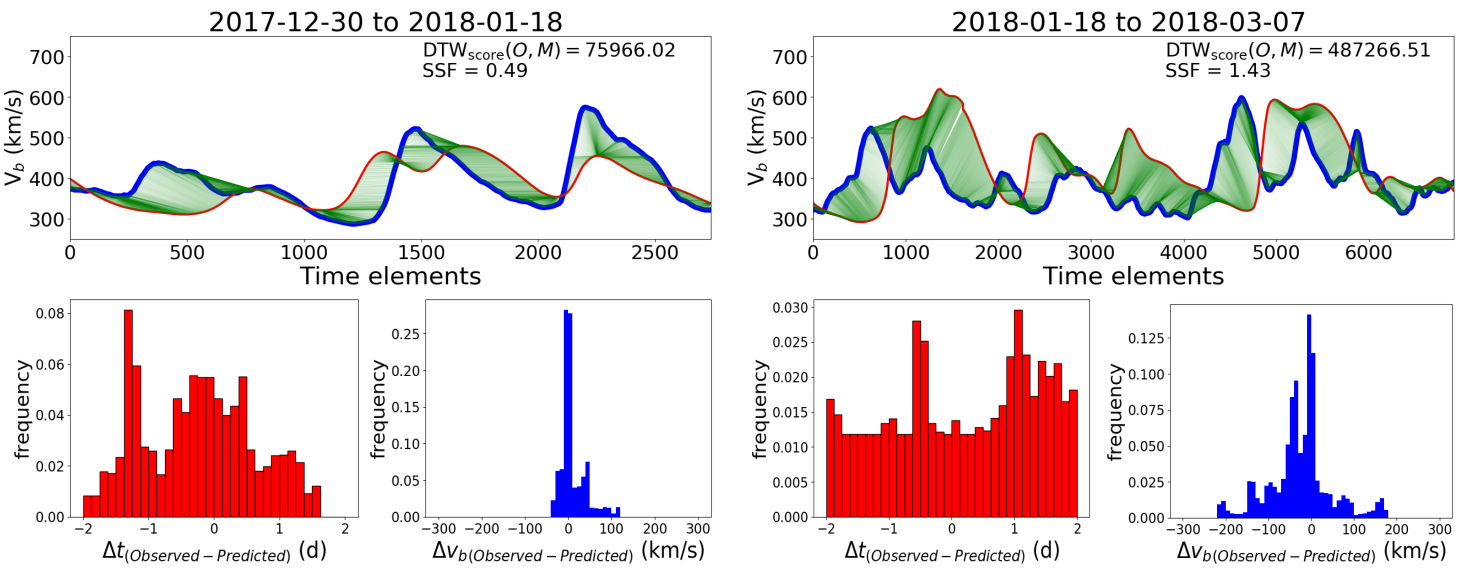}{0.95\textwidth}{}}
\caption{DTW alignments and histograms of time and velocity differences for the first four periods under assessment between November 2017 and September 2019.}
\label{Fig:Appendix_plots1}
\end{figure}

\begin{figure}
\centering
\gridline{\fig{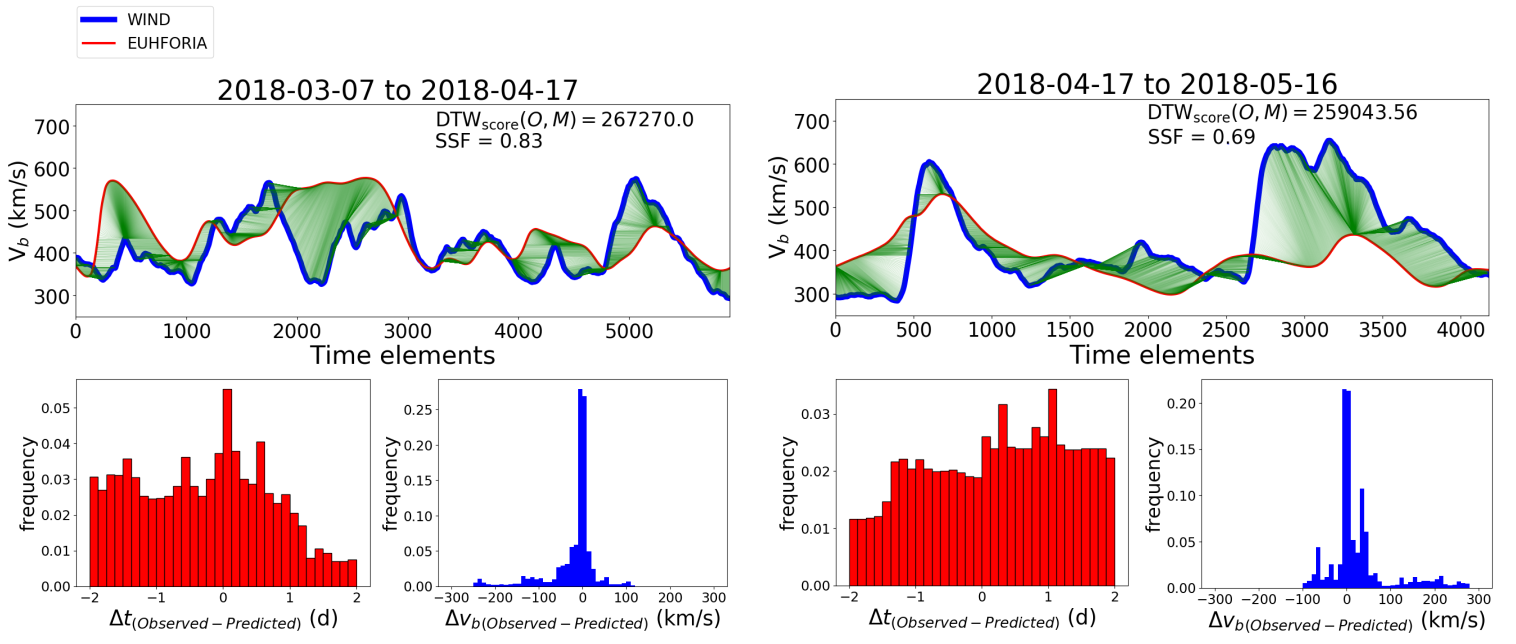}{0.95\textwidth}{}}
            
\gridline{\fig{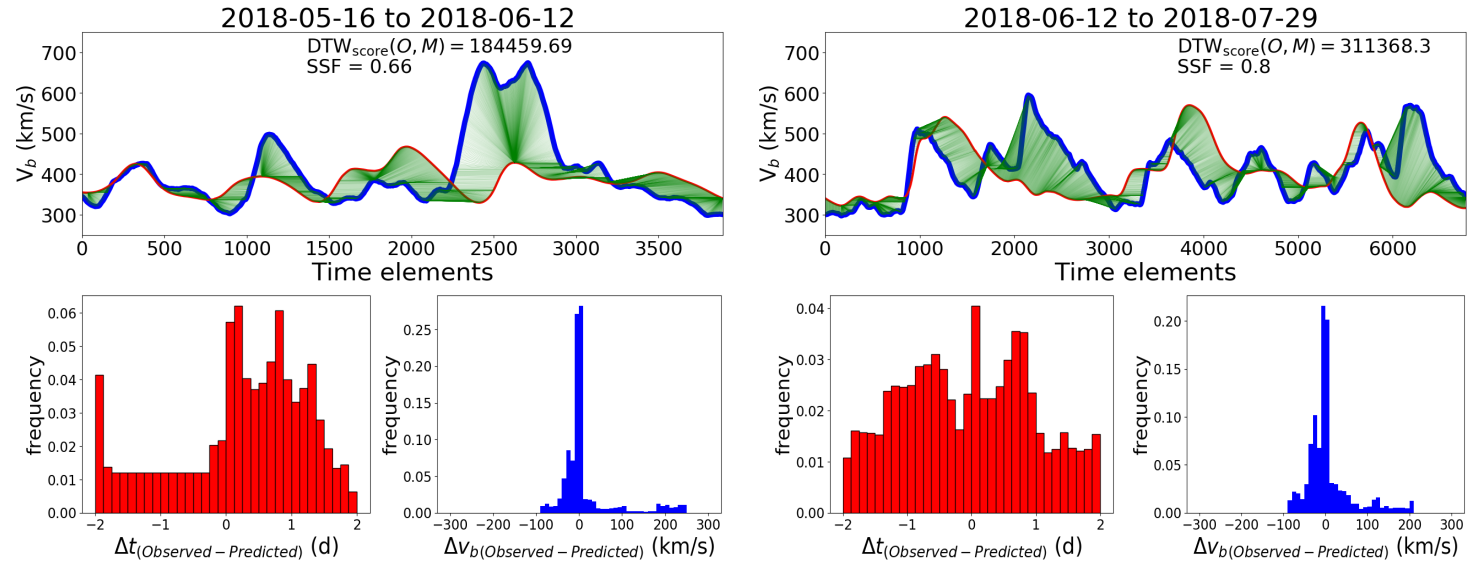}{0.95\textwidth}{}}

\gridline{\fig{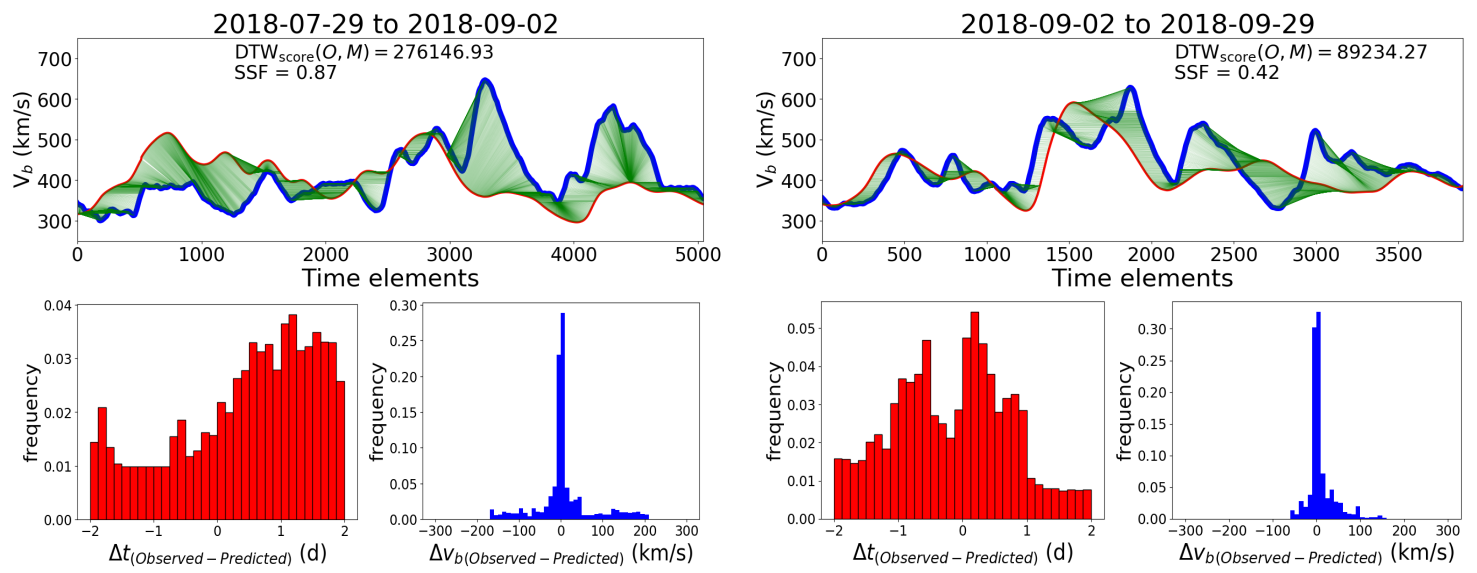}{0.95\textwidth}{}}
           
\caption{DTW alignments and histograms of time and velocity differences for the next six periods under assessment between November 2017 and September 2019 (continuation from Fig.~\ref{Fig:Appendix_plots1}).}
\label{Fig:Appendix_plots2}
\end{figure}

\begin{figure}
\centering
\gridline{\fig{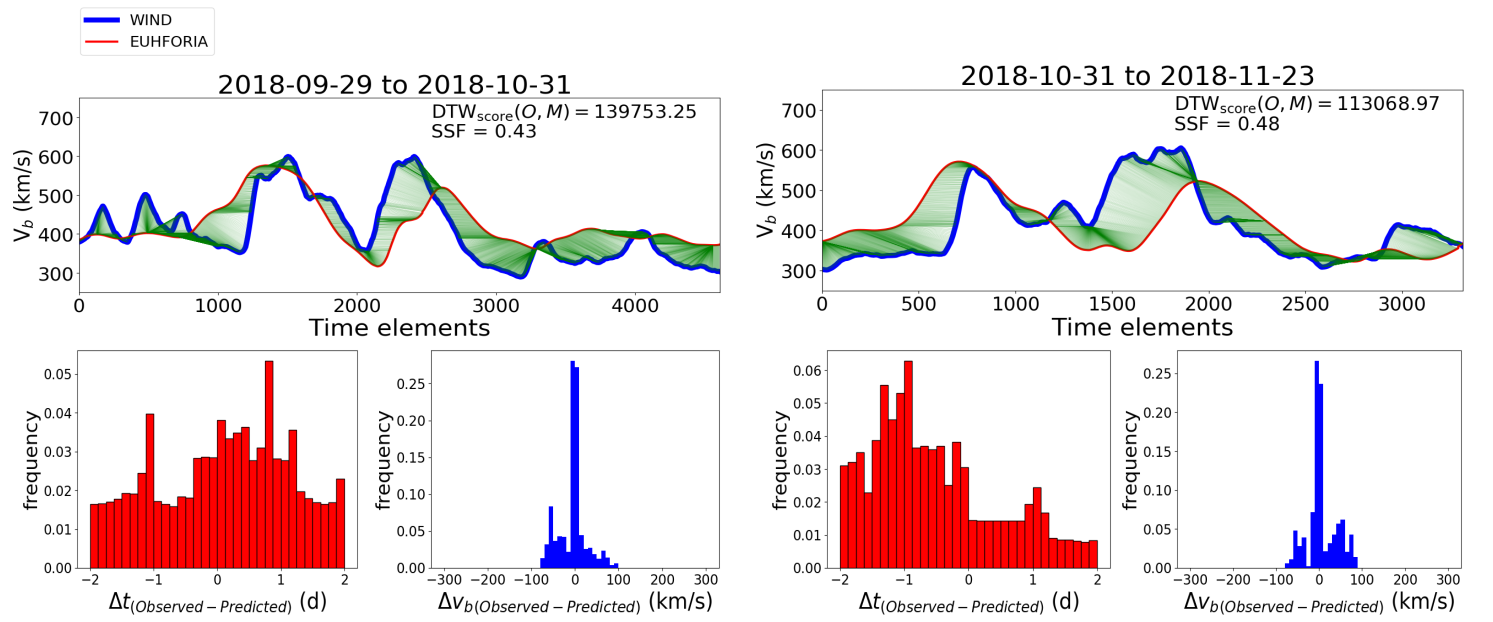}{0.95\textwidth}{}}
            
\gridline{\fig{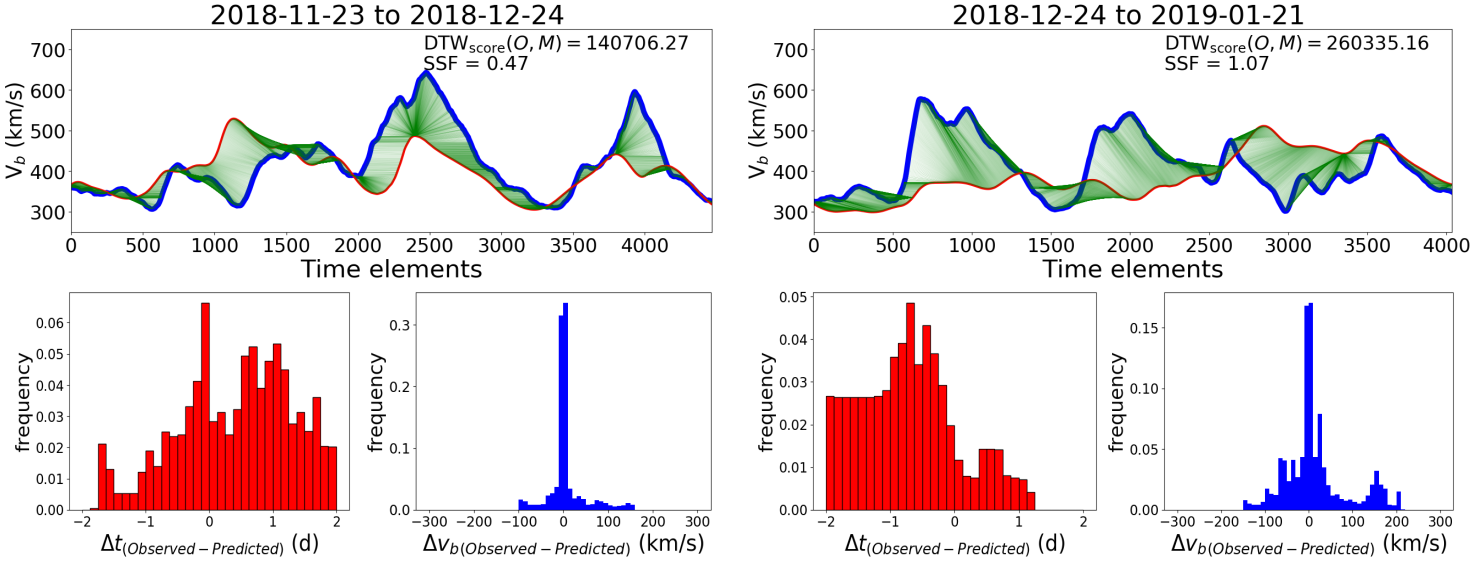}{0.95\textwidth}{}}

\gridline{\fig{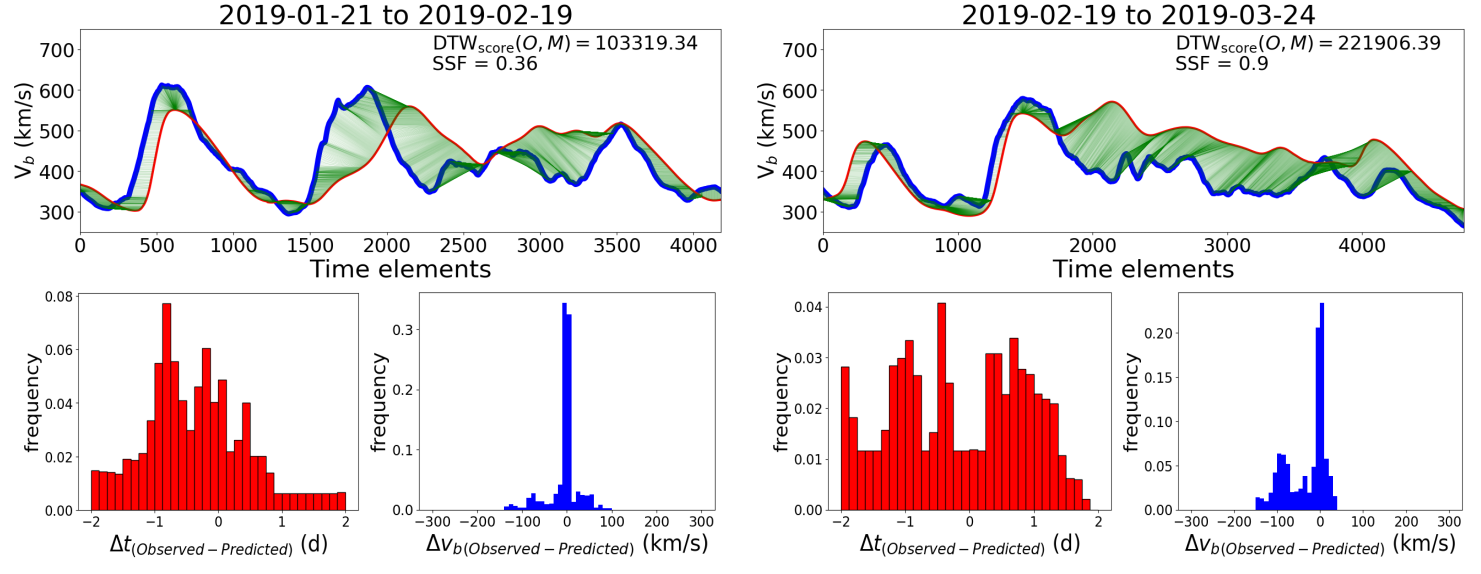}{0.95\textwidth}{}}
           
\caption{DTW alignments and histograms of time and velocity differences for the next six periods under assessment between November 2017 and September 2019 (continuation from Fig.~\ref{Fig:Appendix_plots2}).}
\label{Fig:Appendix_plots3}
\end{figure}

\begin{figure}
\centering
\gridline{\fig{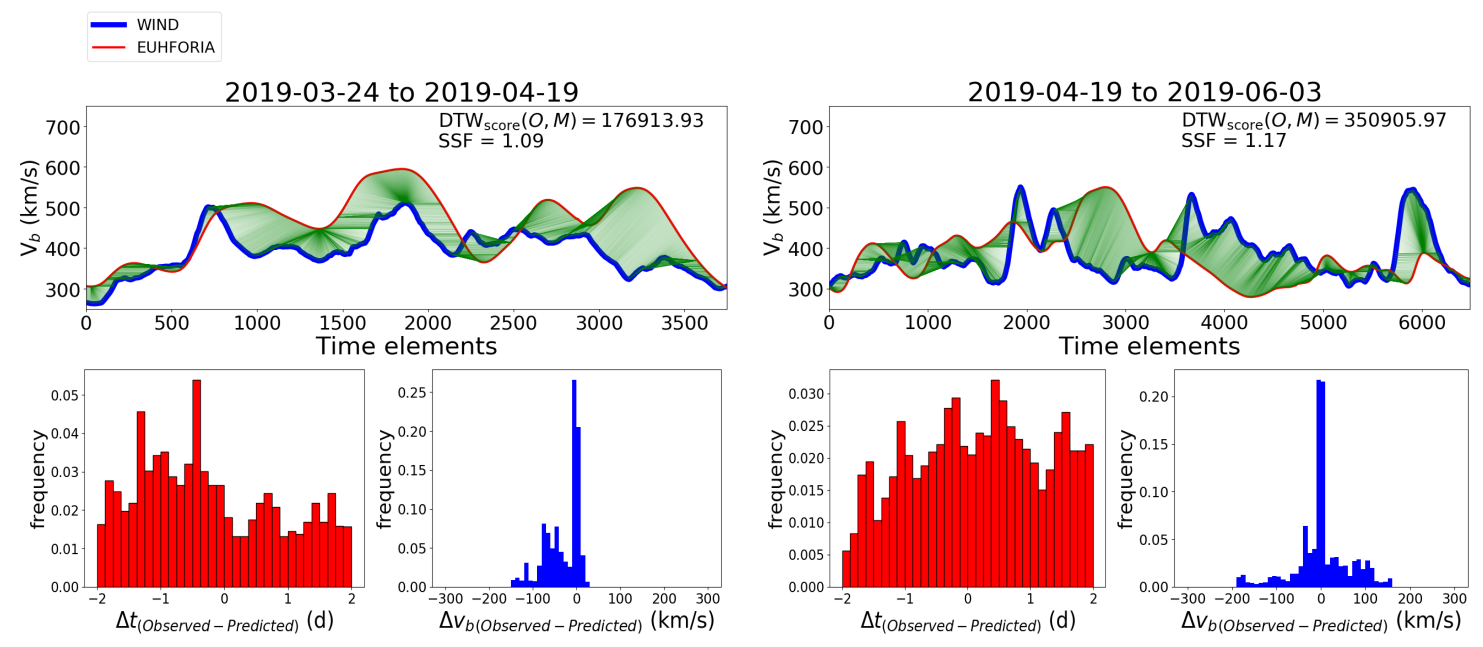}{0.95\textwidth}{}}
            
\gridline{\fig{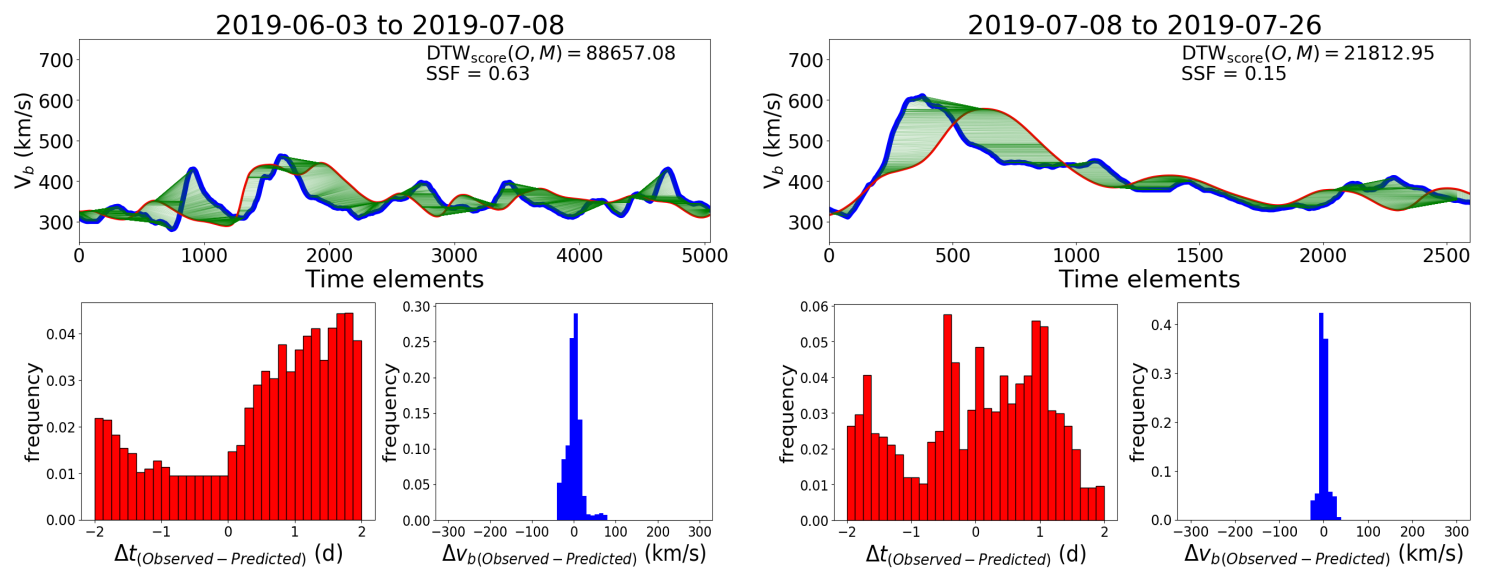}{0.95\textwidth}{}}

\gridline{\fig{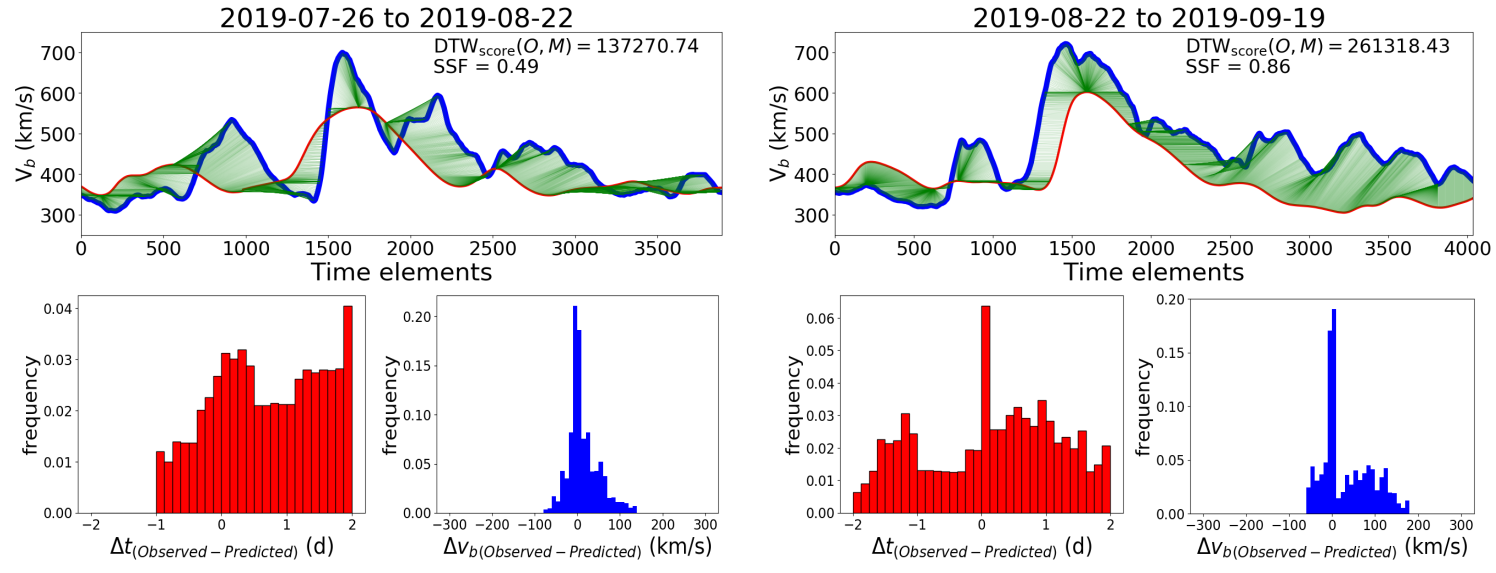}{0.95\textwidth}{}}
           
\caption{DTW alignments and histograms of time and velocity differences for the next six periods under assessment between November 2017 and September 2019 (continuation from Fig.~\ref{Fig:Appendix_plots3}).}
\label{Fig:Appendix_plots4}
\end{figure}

In Fig.~\ref{Fig:Appendix_plots1}-\ref{Fig:Appendix_plots4}, we present the optimal DTW alignments between WIND observations and EUHFORIA predictions as well as the histograms of time and velocity differences for all the individual periods listed in Table \ref{Table:DTWTable}. Green lines show how points from the blue time series are matched with points from the red time series. The DTW score as well as the SSF are shown in the upper right part of each time series plot. Moreover, the histograms of each period provide a good idea of the minimum and maximum differences in time and velocities between the two sequences.

\clearpage

\begin{acknowledgments}
The authors would like to acknowledge the anonymous referee for the fruitful comments which improved the content of the paper. They would also like to thank Emmanuel Chan$\textrm{\'e}$ for his big contribution and insight in this work which helped the project grow. Moreover, they extend their acknowledgments to Eamonn Keogh for the helpful discussions and advice on DTW. E.S. was supported by PhD grants awarded by the Royal Observatory of Belgium. C.V. is funded by the Research Foundation – Flanders, FWO SB PhD fellowship 11ZZ216N. EUHFORIA is developed as a joint effort between the KU Leuven and the University of Helsinki. Work at IRAP was supported by the Centre National de la Recherche Scientifique (CNRS, France), the Centre National d’Etudes Spatiales (CNES, France), the Université Paul Sabatier (UPS). The validation of solar wind with EUHFORIA is being performed within the BRAIN-be project SWiM (Solar Wind Modeling with EUHFORIA for the new heliospheric missions). This project has received funding from the European Union’s Horizon 2020 research and innovation programs under grant agreements No 870405 (EUHFORIA 2.0) and 870437 (SafeSpace). SP acknowledges support from the projects C14/19/089  (C1 project Internal Funds KU Leuven), G.0D07.19N  (FWO-Vlaanderen), and SIDC Data Exploitation (ESA Prodex-12). Computational resources and services used in this work were provided by the VSC (Flemish Supercomputer Center), funded by the Research Foundation Flanders (FWO) and the Flemish Government-Department EWI. 
\end{acknowledgments}

\software{The DTW code used for this work was based on \citet{DTW_AlgorithmReview, 1stNiceDTWarticle, 2ndNiceDTWarticle} and can be found at \url{https://github.com/SamaraEvangelia/DTW_ForSolarWindEvaluation}}.

\bibliography{bibliography}{}
\bibliographystyle{aasjournal}



\end{document}